\documentclass[12pt]{article}
\usepackage[dvips]{graphicx}
\usepackage{float}
\usepackage{epsfig}
\usepackage{ulem}
\usepackage{latexsym,amsmath,amsfonts,amssymb}
\usepackage[latin1]{inputenc}
\usepackage{rotating}
\usepackage[american]{babel}
\usepackage[dvips]{graphicx}
\usepackage{bbm}
\usepackage{color}
\usepackage{slashed}
\usepackage[unicode]{hyperref}
\usepackage{lscape}
\usepackage{bigints}
\def\im{Invent. Math.}

\def\hat{\widehat}
\def\a{\alpha}
\def\b{\beta}
\def\c{\gamma}
\def\d{\delta}
\def\f{\phi}               
\def\vf{\varphi}  
\def\tvf{\tilde{\varphi}}
\def\vp{\varphi}
\def\g{\gamma}
\def\h{\eta}
\def\j{\psi}
\def\k{\kappa}                    
\def\l{\lambda}
\def\m{\mu}
\def\n{\nu}
\def\o{\omega}  \def\w{\omega}

\def\q{\theta}  \def\th{\theta}                  
\def\r{\rho}                                     
\def\s{\sigma}                                   
\def\t{\tau}
\def\u{\upsilon}
\def\x{\xi}
\def\z{\zeta}
\def\pt{\tilde{\varphi}}
\def\tt{\tilde{\theta}}
\def\lab{\label}
\def\6{\partial}
\def\wg{\wedge}
\def\bpsi{\bar{\psi}}
\def\bt{\bar{\theta}}
\def\bvf{\bar{\varphi}}

\DeclareMathOperator{\tr}{tr}

\newcommand{\be}{\begin{equation}}
\newcommand{\ee}{\end{equation}}
\newcommand{\beq}{\begin{equation}}
\newcommand{\eeq}{\end{equation}}
\newcommand{\bea}{\begin{eqnarray}}
\newcommand{\eea}{\end{eqnarray}}
\newcommand{\nn}{\nonumber}

\newcommand{\ba}{\begin{eqnarray}}
\newcommand{\ea}{\end{eqnarray}}

\newcommand{\beqs}{\begin{eqnarray}}
\newcommand{\eeqs}{\end{eqnarray}}
\newcommand{\bal}{\begin{aligned}}
\newcommand{\eal}{\end{aligned}}
\makeatletter
\newcommand\setItemnumber[1]{\setcounter{enum\romannumeral\@enumdepth}{\numexpr#1-1\relax}}
\makeatother

\begin{document}
\baselineskip=15.5pt
\pagestyle{plain}
\setcounter{page}{1}

\def\del{{\partial}}
\def\vev#1{\left\langle #1 \right\rangle}
\def\cn{{\cal N}}
\def\co{{\cal O}}


\def\IC{{\mathbb C}}
\def\IR{{\mathbb R}}
\def\IZ{{\mathbb Z}}
\def\RP{{\bf RP}}
\def\CP{{\bf CP}}
\def\Poincare{{Poincar\'e }}
\def\tr{{\rm tr}}
\def\tp{{\tilde \Phi}}

\def\TL{\hfil$\displaystyle{##}$}
\def\TR{$\displaystyle{{}##}$\hfil}
\def\TC{\hfil$\displaystyle{##}$\hfil}
\def\TT{\hbox{##}}
\def\HLINE{\noalign{\vskip1\jot}\hline\noalign{\vskip1\jot}}
\def\seqalign#1#2{\vcenter{\openup1\jot
   \halign{\strut #1\cr #2 \cr}}}
\def\lbldef#1#2{\expandafter\gdef\csname #1\endcsname {#2}}
\def\eqn#1#2{\lbldef{#1}{(\ref{#1})}%
\begin{equation} #2 \label{#1} \end{equation}}
\def\eqalign#1{\vcenter{\openup1\jot
     \halign{\strut\span\TL & \span\TR\cr #1 \cr
    }}}

\def\eno#1{(\ref{#1})}
\def\href#1#2{#2}
\def\half{\frac{1}{2}}



\def\ads{{\it AdS}}
\def\adsp{{\it AdS}$_{p+2}$}
\def\cft{{\it CFT}}

\newcommand{\ber}{\begin{eqnarray}}
\newcommand{\eer}{\end{eqnarray}}

\newcommand{\beqar}{\begin{eqnarray}}
\newcommand{\cN}{{\cal N}}
\newcommand{\cO}{{\cal O}}
\newcommand{\cA}{{\cal A}}
\newcommand{\cT}{{\cal T}}
\newcommand{\cF}{{\cal F}}
\newcommand{\cC}{{\cal C}}
\newcommand{\cR}{{\cal R}}
\newcommand{\cW}{{\cal W}}
\newcommand{\eeqar}{\end{eqnarray}}
\newcommand{\tht}{\thteta}
\newcommand{\lm}{\lambda}\newcommand{\Lm}{\Lambda}


\newcommand{\nonu}{\nonumber}
\newcommand{\oh}{\displaystyle{\frac{1}{2}}}
\newcommand{\dsl}
   {\kern.06em\hbox{\raise.15ex\hbox{$/$}\kern-.56em\hbox{$\partial$}}}
\newcommand{\id}{i\!\!\not\!\partial}
\newcommand{\as}{\not\!\! A}
\newcommand{\ps}{\not\! p}
\newcommand{\ks}{\not\! k}
\newcommand{\D}{{\cal{D}}}
\newcommand{\dv}{d^2x}
\newcommand{\Z}{{\cal Z}}
\newcommand{\N}{{\cal N}}
\newcommand{\Dsl}{\not\!\! D}
\newcommand{\Bsl}{\not\!\! B}
\newcommand{\Psl}{\not\!\! P}

\newcommand{\eeqarr}{\end{eqnarray}}
\newcommand{\ZZ}{{\rm \kern 0.275em Z \kern -0.92em Z}\;}


\def\del{{\delta^{\hbox{\sevenrm B}}}} \def\ex{{\hbox{\rm e}}}
\def\azb{A_{\bar z}} \def\az{A_z} \def\bzb{B_{\bar z}} \def\bz{B_z}
\def\czb{C_{\bar z}} \def\cz{C_z} \def\dzb{D_{\bar z}} \def\dz{D_z}
\def\im{{\hbox{\rm Im}}} \def\mod{{\hbox{\rm mod}}} \def\tr{{\hbox{\rm Tr}}}
\def\ch{{\hbox{\rm ch}}} \def\imp{{\hbox{\sevenrm Im}}}
\def\trp{{\hbox{\sevenrm Tr}}} \def\vol{{\hbox{\rm Vol}}}
\def\rl{\Lambda_{\hbox{\sevenrm R}}} \def\wl{\Lambda_{\hbox{\sevenrm W}}}
\def\fc{{\cal F}_{k+\cox}} \def\vev{vacuum expectation value}
\def\nodiv{\mid{\hbox{\hskip-7.8pt/}}}
\def\ie{{\em i.e.}}
\def\ie{\hbox{\it i.e.}}

\def\CC{{\mathchoice
{\rm C\mkern-8mu\vrule height1.45ex depth-.05ex
width.05em\mkern9mu\kern-.05em}
{\rm C\mkern-8mu\vrule height1.45ex depth-.05ex
width.05em\mkern9mu\kern-.05em}
{\rm C\mkern-8mu\vrule height1ex depth-.07ex
width.035em\mkern9mu\kern-.035em}
{\rm C\mkern-8mu\vrule height.65ex depth-.1ex
width.025em\mkern8mu\kern-.025em}}}

\def\RR{{\rm I\kern-1.6pt {\rm R}}}
\def\NN{{\rm I\!N}}
\def\ZZ{{\rm Z}\kern-3.8pt {\rm Z} \kern2pt}
\def\IB{\relax{\rm I\kern-.18em B}}
\def\ID{\relax{\rm I\kern-.18em D}}
\def\II{\relax{\rm I\kern-.18em I}}
\def\IP{\relax{\rm I\kern-.18em P}}
\newcommand{\CS}{{\scriptstyle {\rm CS}}}
\newcommand{\CSs}{{\scriptscriptstyle {\rm CS}}}
\newcommand{\rc}{\nonumber\\}
\newcommand{\bear}{\begin{eqnarray}}
\newcommand{\eear}{\end{eqnarray}}

\newcommand{\LL}{{\cal L}}

\def\mani{{\cal M}}
\def\calo{{\cal O}}
\def\calb{{\cal B}}
\def\calw{{\cal W}}
\def\calz{{\cal Z}}
\def\cald{{\cal D}}
\def\calc{{\cal C}}

\def\to{\rightarrow}
\def\ele{{\hbox{\sevenrm L}}}
\def\ere{{\hbox{\sevenrm R}}}
\def\zb{{\bar z}}
\def\wb{{\bar w}}
\def\nodiv{\mid{\hbox{\hskip-7.8pt/}}}
\def\menos{\hbox{\hskip-2.9pt}}
\def\dr{\dot R_}
\def\drr{\dot r_}
\def\ds{\dot s_}
\def\da{\dot A_}
\def\dga{\dot \gamma_}
\def\ga{\gamma_}
\def\dal{\dot\alpha_}
\def\al{\alpha_}
\def\cl{{closed}}
\def\cls{{closing}}
\def\vev{vacuum expectation value}
\def\tr{{\rm Tr}}
\def\to{\rightarrow}
\def\too{\longrightarrow}


\def\a{\alpha}
\def\b{\beta}
\def\c{\gamma}
\def\d{\delta}
\def\e{\epsilon}           
\def\F{\Phi}
\def\f{\phi}               
\def\vf{\varphi}  \def\tvf{\tilde{\varphi}}
\def\vp{\varphi}
\def\g{\gamma}
\def\h{\eta}
\def\j{\psi}
\def\k{\kappa}                    
\def\l{\lambda}
\def\m{\mu}
\def\n{\nu}
\def\o{\omega}  \def\w{\omega}
\def\q{\theta}  \def\th{\theta}                  
\def\r{\rho}                                     
\def\s{\sigma}                                   
\def\t{\tau}
\def\u{\upsilon}
\def\x{\xi}
\def\X{\Xi}
\def\z{\zeta}
\def\pt{\tilde{\varphi}}
\def\tt{\tilde{\theta}}
\def\lab{\label}
\def\6{\partial}
\def\wg{\wedge}
\def\atanh{{\rm arctanh}}
\def\bpsi{\bar{\psi}}
\def\bt{\bar{\theta}}
\def\bvf{\bar{\varphi}}

%



\newfont{\namefont}{cmr10}
\newfont{\addfont}{cmti7 scaled 1440}
\newfont{\boldmathfont}{cmbx10}
\newfont{\headfontb}{cmbx10 scaled 1728}





\newcommand{\re}{\,\mathbb{R}\mbox{e}\,}
\newcommand{\hyph}[1]{$#1$\nobreakdash-\hspace{0pt}}
\providecommand{\abs}[1]{\lvert#1\rvert}
\newcommand{\Nugual}[1]{$\mathcal{N}= #1 $}
\newcommand{\sub}[2]{#1_\text{#2}}
\newcommand{\partfrac}[2]{\frac{\partial #1}{\partial #2}}
\newcommand{\bsp}[1]{\begin{equation} \begin{split} #1 \end{split} \end{equation}}
\newcommand{\calF}{\mathcal{F}}
\newcommand{\calO}{\mathcal{O}}
\newcommand{\calM}{\mathcal{M}}
\newcommand{\calV}{\mathcal{V}}
\newcommand{\bbZ}{\mathbb{Z}}
\newcommand{\bbC}{\mathbb{C}}
\newcommand{\cK}{{\cal K}}

\newcommand{\Thq}{\Theta\left(\r-\r_q\right)}
\newcommand{\Dq}{\d\left(\r-\r_q\right)}
\newcommand{\kten}{\kappa^2_{\left(10\right)}}
\newcommand{\pbi}[1]{\imath^*\left(#1\right)}
\newcommand{\ho}{\hat{\omega}}
\newcommand{\tth}{\tilde{\th}}
\newcommand{\tf}{\tilde{\f}}
\newcommand{\tj}{\tilde{\j}}
\newcommand{\tw}{\tilde{\omega}}
\newcommand{\tz}{\tilde{z}}
\newcommand{\prj}[2]{(\partial_r{#1})(\partial_{\j}{#2})-(\partial_r{#2})(\partial_{\j}{#1})}
\def\atanh{{\rm arctanh}}
\def\sech{{\rm sech}}
\def\csch{{\rm csch}}
\allowdisplaybreaks[1]

\def\red{\textcolor[rgb]{0.98,0.00,0.00}}

\newcommand{\Dan}[1] {{\textcolor{blue}{#1}}}

\numberwithin{equation}{section}

\newcommand{\Tr}{\mbox{Tr}}    


%

\setcounter{footnote}{0}
\renewcommand{\theequation}{{\rm\thesection.\arabic{equation}}}

\begin{titlepage}

\begin{center}

\vskip .5in 
\noindent

{\Large \bf{ AdS$_2$ duals to ADHM quivers with Wilson lines} }
\bigskip\medskip

Yolanda Lozano$^{a,}$\footnote{ylozano@uniovi.es}, Carlos Nunez$^{b,}$\footnote{c.nunez@swansea.ac.uk}, Anayeli Ramirez$^{a,}$\footnote{anayelam@gmail.com} and Stefano Speziali$^{b,}$\footnote{stefano.speziali6@gmail.com}\\

\bigskip\medskip
{\small

 $a$: Department of Physics, University of Oviedo,
Avda. Federico Garcia Lorca s/n, 33007 Oviedo, Spain
\vskip 3mm
 $b$: Department of Physics, Swansea University, Swansea SA2 8PP, United Kingdom}

\vskip .5cm 
\vskip .9cm 
     	{\bf Abstract }

\vskip .1in
\end{center}

\noindent
We discuss  AdS$_2\times S^3\times{\text{CY}}_2\times I_{\rho}$ solutions to massive Type IIA supergravity with 4 Poincar\'e supersymmetries. We propose explicit dual quiver quantum mechanics built out of D0 and D4 colour branes coupled to D4' and D8 flavour branes. We propose that these quivers describe the interactions of instantons and Wilson lines in 5d gauge theories with 8 Poincar\'e supersymmetries. Using the RR Maxwell fluxes of the solutions, { conveniently put off-shell, we construct a functional 
from which the holographic central charge can be derived through a geometrical extremisation principle}. %
\noindent
\vskip .5cm
\vskip .5cm
\vfill
\eject

\end{titlepage}

\setcounter{footnote}{0}

\tableofcontents
\newpage
\renewcommand{\theequation}{{\rm\thesection.\arabic{equation}}}
 \section{Introduction}

Recent progress in Holography deepened our understanding of lower dimensional realisations of the AdS/CFT correspondence \cite{Maldacena:1997re}. These scenarios are particularly relevant for the development of the black hole microstate counting programme. Indeed, 4d and 5d extremal black holes exhibit AdS$_2$ and AdS$_3$ geometries close to their horizons. 

Efforts in classifying AdS$_2$ and AdS$_3$ geometries in M-theory and Type II supergravity have revealed that a plethora of these solutions exists, exhibiting different geometrical structures and preserving different amounts of supersymmetries, see for example \cite{Argurio:2000tg}-\cite{Lozano:2020txg}.
 Indeed, as the dimensionality of the internal space increases, the richer the structure of all possible geometries and topologies thereof becomes. Exhausting all possibilities through classifications becomes then increasingly complex.

Major recent progress has been achieved in the classification of AdS$_3$ geometries with $\mathcal{N}=(0,4)$ supersymmetries and SU(2) structure \cite{Couzens:2017way,Lozano:2019emq}. Remarkably, the 2d CFTs dual to subsets of these solutions have been identified   \cite{Lozano:2019zvg,Lozano:2019jza,Lozano:2019ywa} (see also \cite{Filippas:2019ihy}), thus  providing for explicit AdS$_3$/CFT$_2$ pairs where the microscopical description of 5d black holes can be addressed. Generalising recent developments in five dimensional BPS black holes (see for instance  \cite{Haghighat:2015ega}-\cite{Hull:2020byc})  to these new set-ups constitutes a promising new avenue that is still awaiting for further development.

AdS$_2$ geometries arise as near horizon geometries of 4d extremal black holes, and are thus ubiquitous in their microscopical studies. The precise realisation of the AdS$_2$/CFT$_1$ holographic correspondence presents however important technical and conceptual problems \cite{Maldacena:1998uz,Denef:2007yt,Maldacena:2016hyu,Maldacena:2016upp}. These mainly originate from the fact that the boundary of AdS$_2$ is non-connected \cite{Harlow:2018tqv}. As a result this correspondence is much less understood than its higher dimensional counterparts.

A promising approach to the study of the AdS$_2$/CFT$_1$ correspondence is to exploit its connection with the better understood AdS$_3$/CFT$_2$ correspondence. This approach has been explored recently in \cite{Lozano:2020txg}. At the geometrical level the AdS$_3$ and AdS$_2$ spaces are related by Abelian T-duality.  
At the level of the dual CFTs the Superconformal Quantum Mechanics (SCQM) dual to the AdS$_2$ solutions arise from the 2d CFTs dual to the AdS$_3$ backgrounds upon dimensional reduction. In this manner
new families of AdS$_2$ solutions to Type IIB supergravity with $\mathcal{N}=4$ supercharges and their dual SCQMs were constructed in \cite{Lozano:2020txg}, using as seed solutions the $\mathcal{N}=(0,4)$ AdS$_3$ solutions to massive Type IIA supergravity recently found in \cite{Lozano:2020bxo}. 
These constructions provide explicit string theory set-ups in which one of the chiral sectors of the 2d CFT is decoupled in the SCQM, as explained in \cite{Balasubramanian:2009bg,Azeyanagi:2007bj,Aniceto:2020saj}. Remarkably, the corresponding quivers inherit, by construction, many of the properties of the parent 2d quiver CFTs, like the matter content that guarantees gauge anomaly cancellation in 2d.

Given that SCQMs do not have these constraints, one would expect that more general quivers than those arising upon reduction could be constructed. With this goal in mind we will follow in this paper an alternative road to the study of AdS$_2$/CFT$_1$ pairs. Our starting point will be the 
new class of AdS$_2$ solutions with $\mathcal{N}=4$ supercharges recently constructed in \cite{Lozano:2020bxo}, using the technique of double analytical continuation. Therefore, a priori one does not expect any relation between these solutions and 2d CFTs\footnote{Besides the fact that, as we will see, the 1d dual CFTs will be formulated in terms of (0,4) 2d matter fields.}. 

The technique of double analytical continuation has been extensively used in the construction of AdS$_2$ solutions. Indeed, the only requirement to produce an AdS$_2$ solution is that an $S^2$ exists in the internal space of an already known supergravity background, that can be analytically continued to AdS$_2$. Fortunately, many AdS$_p$ supergravity solutions contain $S^2$'s in their transverse spaces.  The AdS$_p$ subspace itself gives rise upon the analytical continuation to an $S^p$ that is part of the internal space of the new solutions, and realises some of its isometries.
Some examples of AdS$_2$ solutions constructed in this way are the AdS$_2\times S^6$ backgrounds constructed  in \cite{Corbino:2017tfl,Corbino:2018fwb} from the class of AdS$_6\times S^2$ solutions to Type IIB in 
\cite{DHoker:2016ujz}, the AdS$_2\times S^7$ solutions to massive Type IIA constructed in \cite{Dibitetto:2018gbk} from the AdS$_7\times S^2$ solutions of \cite{Apruzzi:2013yva}, or the AdS$_2\times S^4\times S^2$ solutions constructed in \cite{Dibitetto:2019nyz}, from the AdS$_4\times S^2\times S^2$ backgrounds in \cite{DHoker:2007zhm}. These solutions preserve different amounts of supersymmetries, and allow, in some cases, for interesting interpretations as line defect CFTs \cite{Dibitetto:2018gtk,Chen:2020mtv,Faedo:2020nol} or deconstructed higher dimensional CFTs \cite{Dibitetto:2019nyz}. 

In this paper we focus our study on the AdS$_2\times S^3\times \text{CY}_2\times I_\rho$ solutions to massive Type IIA supergravity recently constructed in \cite{Lozano:2020bxo}. These solutions can be thought of as the dual description of the quantum mechanics of a point like defect in a five-dimensional CFT. The contents of this work are distributed as follows.
In section 2 we review the main properties of these solutions, we compute the quantised charges and propose the underlying brane set-up. This is a 1/8 BPS D0-D4-D4'-D8-F1 brane intersection, previously studied in \cite{Dibitetto:2018gtk,Lozano:2020bxo,Faedo:2020nol}. We show that the D0, D4 and F1 branes are most naturally counted with their (regularised) electric charges, rather than their magnetic ones. This is in agreement with their interpretation in terms of instantons in the worldvolumes of the D4' and D8 branes, interacting with Wilson lines. In section 3 we set to discuss the superconformal mechanics dual to these solutions. We start reviewing the CFT dual to the D0-D4-F1 system, described in the near horizon limit by one of our solutions. This example is used to introduce the low-energy fields that will enter in the quantum mechanics dual to more general solutions. We construct explicit quiver quantum mechanics that we interpret as describing D0 and D4 brane instantons in the worldvolumes of D4' and D8 branes, interacting with Wilson lines in the antisymmetric representations of their gauge groups. We use the notion of central charge  for superconformal quantum mechanics put forward in \cite{Lozano:2020txg}, defined from the number of vacuum states of the theory, and compare it to the holographic calculation. In section 4 we further elaborate on the relation between the holographic central charge and the product of electric and magnetic RR charges of AdS$_2$ solutions pointed out in \cite{Lozano:2020txg}. {Furthermore, we propose an extremising functional constructed from the RR Maxwell fluxes from which  we derive the holographic central charge through an extremisation principle}. Section 5 contains our conclusions and future directions. Appendix A contains a summary of the AdS$_3\times S^2\times \text{CY}_2$ solutions and their 2d dual CFTs studied in  \cite{Lozano:2019emq,Lozano:2019zvg,Lozano:2019jza,Lozano:2019ywa}. These motivate the families of backgrounds for which we construct dual SCQMs in this paper. Appendix B contains a detailed derivation of the low-energy field content of the D0-D4-D4'-D8 brane web. Appendix C studies different probe branes of interest in our backgrounds. Finally, Appendices D and E contain specific geometrical properties of our solutions.

\section{The AdS$_2\times S^3\times$ CY$_2$ solutions to massive IIA}\label{seccionC}

In this section we present and further discuss the AdS$_2\times S^3\times$ CY$_2$ solutions to massive IIA supergravity constructed in \cite{Lozano:2020bxo}. These solutions were obtained
through a double analytic continuation of the AdS$_3\times S^2\times$ CY$_2$ backgrounds to massive IIA constructed in \cite{Lozano:2019emq}, and summarised in Appendix \ref{AdS3} (see eqs. \eqref{kkktmba}, \eqref{eq:classIflux2} and \eqref{eq:classIflux3}). We give a thorough study of the underlying geometry that will allow us to propose concrete superconformal quantum mechanics dual to the solutions. 

The class of solutions referred as class I in \cite{Lozano:2020bxo} contain a transverse space M$_4=$CY$_2$\footnote{A second class of solutions, referred as class II, contain an M$_4$  K\"ahler manifold.}. They are given by
\begin{equation}\label{NS sector analytically continued}
\begin{split}
\text{d}s^2 &= \frac{u}{\sqrt{\hat{h}_4 h_8}} \left( \frac{\hat{h}_4 h_8}{4 \hat{h}_4 h_8 - (u')^2} \text{d}s^2_{\text{AdS}_2} + \text{d}s^2_{S^3}\right) + \sqrt{\frac{\hat{h}_4}{h_8}} \text{d}s^2_{\text{CY}_2} + \frac{\sqrt{\hat{h}_4 h_8}}{u} \text{d} \rho^2 \, , \\
e^{- \Phi}&= \frac{h_8^{3/4}}{2 \hat{h}_4^{1/4} \sqrt{u}} \sqrt{4 \hat{h}_4 h_8 - (u')^2} \, , \quad H_{(3)} = - \frac{1}{2} \text{d} \bigg( \rho + \frac{u u'}{4 \hat{h}_4 h_8 - (u')^2} \bigg) \wedge \hat{\text{vol}}_{\text{AdS$_2$}} + \frac{1}{h_8^2} \text{d} \rho \wedge H_2 \, .
\end{split}
\end{equation}
Here $\Phi$ is the dilaton, $H_{(3)}= \text{d} B_{(2)}$ is the NS 3-form and the metric is written in string frame.
The warping function $\hat{h}_4$ has support on $(\rho,\text{CY}_2)$. On the other hand,  $u$ and $h_8$ only depend of $\rho$. We denote $u'= \partial_{\rho}u$ and similarly for $h_8'$. 
Note that it must be that $4 \hat{h}_4 h_8 - (u')^2 >0$, in order for the metric to be of the correct signature and the dilaton to be real. 

The RR sector reads
\begin{equation}\label{RR sector analytically continued}
\begin{split}
F_{(0)} &= h_8' \, , \quad F_{(2)} = - \frac{1}{h_8}H_2 - \frac{1}{2} \Big( h_8 + \frac{h_8' u' u}{4 h_8 \hat{h}_4 - (u')^2} \Big) \hat{\text{vol}}_{\text{AdS$_2$}} \, , \\
F_{(4)} &= \left( - \text{d} \bigg( \frac{u'u}{2 \hat{h}_4} \bigg) + 2 h_8 \text{d} \rho \right) \wedge \hat{\text{vol}}_{S^3}  - \frac{h_8}{u} \hat{\star}_4 \text{d}_4 h_4 \wedge d \rho - \partial_{\rho} \hat{h}_4 \hat{\text{vol}}_{\text{CY$_2$}} + \\
&+\frac{u' u}{2h_8(4 \hat{h}_4 h_8 - (u')^2)} H_2 \wedge \hat{\text{vol}}_{\text{AdS$_2$}} \, ,
\end{split}
\end{equation}
with the higher dimensional fluxes related to these as $F_{(6)}=-\star_{10} F_{(4)},~F_{(8)}=\star_{10} F_{(2)},~F_{(10)}=-\star_{10} F_{(0)}$.
Supersymmetry holds whenever
\beq \label{susyC}
u''=0,~~~~ H_2+ \hat{\star}_4 H_2=0,
\eeq
where $\hat{\star}_4$ is the Hodge dual on CY$_2$. In what follows we will restrict ourselves to the set of solutions for which $H_2=0$
and
 $\hat{h}_4=\hat{h}_4(\rho)$.
In that case the background is a solution of the massive IIA equations of motion if the functions $\hat{h}_4,h_8$ satisfy the conditions 
(away from localised sources),
\begin{equation}
\hat{h}_4''(\rho)=0,\;\;\;\; h_8''(\rho)=0,\;\;\;\; \label{eqsmotion}
\end{equation}
which make them linear functions of $\rho$.

\subsection{Brane set-up} \label{set-up}

The brane set-up associated to the previous solutions was identified in \cite{Faedo:2020nol} (see also \cite{Dibitetto:2018gtk}), for the case of $u=$ constant\footnote{The $u$ non-constant case was recently analysed in \cite{Dibitetto:2020bsh} for the AdS$_3\times S^2\times \text{CY}_2$ backgrounds from which our solutions are constructed by analytical continuation, in the massless case. The brane intersection involves in this case dyonic branes placed at conical singularities.}. It consists on a D0-F1-D4-D4'-D8 brane intersection, depicted in Table \ref{Table brane web set type IIA3},  
\begin{table}[h]
\begin{center}
\begin{tabular}{c|c|c|c|c|c|c|c|c|c|c}
& $x^0$ & $x^1$ & $x^2$ & $x^3$ & $x^4$ & $x^5$ & $x^6$ & $x^7$ & $x^8$ & $x^9$\\
\hline
D0 & $-$ & & & & & $$ & & &$$ &$$  \\
\hline
 D4 &  $-$ &$-$ &$-$  &$-$  &$-$  & & $$ & $$ & $$ &  \\
\hline
D4' & $-$ & $$ & $$ & $$ & $$ & $$ &$-$ &$-$ &$-$ &$-$ \\
\hline
D8 & $-$ & $-$ & $-$ & $-$ & $-$ & & $-$ & $-$ & $-$ &$-$  \\
\hline F1 &$-$ & $$ & $$ & $$ & $$ & $-$& $$ & & & $$ \\
\end{tabular}
\caption{Brane set-up, where $-$ marks the spacetime directions spanned by the various branes. $x^0$ corresponds to the time direction of the ten dimensional spacetime, $x^1, \dots , x^4$ are the coordinates spanned by the CY$_2$, $x^5$ is the direction where the F1-strings are stretched, and $ x^6, x^7, x^8, x^9$ are the coordinates where the SO$(4)$ symmetry is realised.}
\label{Table brane web set type IIA3}
\end{center}
\end{table}
preserving four supersymmetries. This is compatible with the quantised charges obtained from the Page fluxes, as we show below.

The Page fluxes associated to the background given in \eqref{NS sector analytically continued}, \eqref{RR sector analytically continued},  $\hat{F}=F\wedge e^{-B_{(2)}}$, are given by\footnote{These fluxes take into account the effect of the large gauge transformations $B_2\rightarrow B_2+\pi k \hat{\text{vol}}_{\text{AdS}_2}$, for $k=0,1,\dots, P$. These transformations are performed every time a $\rho$-interval $ [2\pi k,2\pi (k+1)]$ is crossed, as explained below.},
\begin{equation}\label{fluxesw}
\begin{split}
\hat{F}_{(0)} & = h'_8 \, , \\
\hat{F}_{(2)} & =- \frac{1}{2} \Bigl(h_8 - (\rho-2\pi k) h'_8\Bigr) \hat{\text{vol}}_{\text{AdS$_2$}} \, , \\
\hat{F}_{(4)} &= -\hat{h}'_4 \hat{\text{vol}}_{\text{CY$_2$}} -\bigg(2h_8+\frac{u'(u\hat{h}_4'-\hat{h}_4u')}{2\hat{h}_4^2}\bigg)  \hat{\text{vol}}_{S^3} \wedge \text{d}\rho \, , \\
\hat{F}_{(6)} &= \frac{1}{2}\Bigl(\hat{h}_4 - (\rho-2\pi k) \hat{h}_4'\Bigr) \hat{\text{vol}}_{\text{AdS$_2$}} \wedge \hat{\text{vol}}_{\text{CY$_2$}} \\
&-\bigg((\rho-2\pi k) h_8-\frac{(u-(\rho-2\pi k)u')(\hat{h}_4u'-u\hat{h}_4')}{4\hat{h}_4^2}\bigg)   \hat{\text{vol}}_{\text{AdS$_2$}} \wedge \hat{\text{vol}}_{S^3} \wedge \mathrm{d} \rho \, , \\
\hat{F}_{(8)} &= \bigg(2\hat{h}_4+\frac{u'(uh_8'-h_8u')}{2h_8^2}\bigg) \hat{\text{vol}}_{\text{CY$_2$}} \wedge \hat{\text{vol}}_{S^3} \wedge \text{d} \rho \, , \\
\hat{F}_{(10)} &=\bigg((\rho-2\pi k) \hat{h}_4-\frac{(u-(\rho-2\pi k)u')(uh_8'-h_8u')}{4h_8^2}\bigg)
\hat{\text{vol}}_{\text{AdS$_2$}} \wedge \hat{\text{vol}}_{S^3} \wedge \hat{\text{vol}}_{\text{CY$_2$}} \wedge \text{d} \rho \, .
\end{split}
\end{equation}
The F1-branes are electrically charged with respect to the NS-NS 3-form. We compute their charges according to 
\begin{equation}
 Q_{\text{F}1}^e=\frac{1}{(2\pi)^2}\int_{\text{AdS}_2\times \text{I}_{\rho} } H_{(3)}\, ,
 \end{equation}
 in units of $\alpha^\prime=g_s=1$.
We regularise the volume of the AdS$_2$ space such that\footnote{This regularisation prescription is based on the analytical continuation that relates the AdS$_2$ space with an $S^2$.}
\begin{equation} \label{regpres}
\text{Vol}_{\text{AdS}_2}=4\pi\, .
\end{equation}
This gives for $H_{(3)}$ as in (\ref{NS sector analytically continued}) (and $H_2=0$),
\begin{equation}
Q_{\text{F}1}^e=\frac{1}{2\pi}\Bigl(\rho-\frac{uu'}{4h_4h_8-u'^2}\Bigr){\Big{\rvert}}_{\rho_{i}}^{\rho_f},
\end{equation}
where this must be computed at both ends of the $I_\rho$-interval.
 From now and for the rest of the paper we will focus our analysis on the $u=$ constant case. In such case
 \begin{equation}
Q_{\text{F}1}^e= \frac{\rho_f-\rho_i}{2\pi}.
\end{equation}
Therefore, there are $k$ F1-strings for $\rho\in [0,2\pi k]$, with one F1-string being created as we move in $\rho$-intervals of length $2\pi$. Enforcing that with our regularisation prescription (\ref{regpres}) $B_2$ lies in the fundamental region,
 \begin{equation}
 \frac{1}{(2\pi)^2}|\int_{\text{AdS}_2}B_{(2)}|\in [0,1]\, ,
 \end{equation}
then implies that a large gauge transformation of parameter $k$ needs to be performed for $\rho\in [2\pi k,2\pi (k+1)]$, such that
 \begin{equation}
 B_{(2)}=-\frac12(\rho-2k\pi)\hat{\text{vol}}_{\text{AdS}_2}.
 \end{equation}
This affects the RR Page fluxes, as
 \begin{equation}
 \hat{F}_{(p)}\rightarrow \hat{F}_{(p)}-k\pi \hat{F}_{(p-2)}\wedge \hat{\text{vol}}_{\text{AdS}_2},
 \end{equation}
which has already been taken into account in the expressions in (\ref{fluxesw}).

As we will see, the D0 and D4 branes will have an interpretation in the dual field theory as instantons. Therefore, we will characterise them  by their electric charges. We use that the electric charge of a Dp-brane is given by
\begin{equation}\label{electric-charges}
Q_{\text{Dp}}^e=\frac{1}{(2\pi)^{p+1}}\int_{\text{AdS}_2\times \Sigma_{p}} \hat{F}_{(p+2)},
\end{equation}
in units of $\alpha^\prime=g_s=1$.
Substituting the electric components of the $\hat{F}_{(2)}$ and $\hat{F}_{(6)}$ fluxes in (\ref{fluxesw}) and regularising the volume of AdS$_2$ as indicated by equation (\ref{regpres}),
we find for the electric charges of the D0 and D4 branes,
\begin{eqnarray} \label{D0D4electric}
Q_{\text{D0}}^{e}&=& h_8 - (\rho-2\pi k)  h'_8\nonumber \\
Q_{\text{D4}}^e&=&h_4 - (\rho-2\pi k) h'_4 \, ,
\end{eqnarray}
where in the last equation we have used that $\hat{h}_4\equiv \Upsilon h_4$ and $\Upsilon \text{Vol}_{\text{CY}_2}=16\pi^4$. 
Both the D0 and the D4 branes play the role of colour branes in the brane set-up, as implied by the fact that both $\mathrm{d} \hat{F}_{(8)}$ and the second component of $\mathrm{d} \hat{F}_{(4)}$ vanish identically. 

The D4' and D8 branes will find an interpretation in the dual field theory as branes where the D0 and D4 brane instantons live. We will characterise them by their magnetic charges. As usual these are computed from 
\begin{equation} \label{magnetic-charges}
Q_{\text{Dp}}^m=\frac{1}{(2\pi)^{7-p}}\int_{\Sigma_{8-p}}\hat{F}_{(8-p)}\, ,
\end{equation}
in units of $\alpha^\prime=g_s=1$. The second component of $\hat{F}_{(4)}$ in (\ref{fluxesw}) is the Hodge-dual of the $\hat{F}_{(6)}$ used in (\ref{D0D4electric}) to compute the electric charge of the D4-branes. In turn, the first component gives rise to a magnetic charge associated to a second type of D4'-branes. This charge reads  
\begin{equation}
Q_{\text{D4}'}^m = \frac{\text{Vol}_{\text{CY}_2}}{(2 \pi)^3} \, \hat{h}_4' = 2\pi h_4'  \, .  \\
\end{equation}
Given that 
\begin{equation}\label{D4 charges}
\mathrm{d} \hat{F}_{(4)} = \hat{h}_4'' \, \mathrm{d} \rho \wedge \hat{\text{vol}}_{\text{CY}_2} \, ,
\end{equation}
these branes provide sources localised in the $\rho$ direction. They are thus flavour  branes. Being localised in $\rho$ and transverse to the CY$_2$, they are naturally seen to wrap AdS$_2 \times S^3$. 
In turn, given that
$\mathrm{d} \hat{F}_{(0)} \neq 0$ if $h_8'' \neq 0$, according to 
\begin{equation}\label{D8 charges}
 \mathrm{d} \hat{F}_{(0)} = h_8'' \mathrm{d} \rho \, ,
\end{equation}
there are also D8 brane sources localised in the $\rho$ direction, also behaving as flavour branes. They are wrapped on AdS$_2 \times S^3 \times \text{CY}_2$. Their magnetic charge is given by
\begin{equation}
Q_{\text{D8}}^m =  2 \pi h_8' .
\end{equation}

\subsection{The local solutions}\label{local solutions}

For $u=$ constant a generic background in our class is defined by the functions $\hat{h}_4$, $h_8$. We will be interested in solutions that in the $\rho\in [2\pi k,2\pi (k+1)]$ interval are of the form
\begin{equation}
\hat{h}_4^{(k)}=\Upsilon\Bigl(\alpha_k+\frac{\beta_k}{2\pi}(\rho-2\pi k)\Bigr)\, , \qquad 
h_8^{(k)}=\Bigl(\mu_k+\frac{\nu_k}{2\pi}(\rho-2\pi k)\Bigr)\, ,
\end{equation}
with the space starting and ending at $\rho=0$ and $\rho=2\pi (P+1)$, respectively, where we take both $\hat{h}_4$ and $h_8$ to vanish. We thus have that
\begin{equation} \label{profileh4sp}
\hat{h}_4(\rho)\!=\!\Upsilon\! \,h_4(\rho)\!=\!\!
                    \Upsilon\!\!\left\{ \begin{array}{ccrcl}
                       \frac{\beta_0 }{2\pi}
                       \rho & 0\leq \rho\leq 2\pi \\
                                     \alpha_k\! +\! \frac{\beta_k}{2\pi}(\rho-2\pi k) &~~ 2\pi k\leq \rho \leq 2\pi(k+1),\;\; k=1,..,P-1\\
                      \alpha_P-  \frac{\alpha_P}{2\pi}(\rho-2\pi P) & 2\pi P\leq \rho \leq 2\pi(P+1),
                                             \end{array}
\right.
\end{equation}
 \begin{equation} \label{profileh8sp}
h_8(\rho)
                    =\left\{ \begin{array}{ccrcl}
                       \frac{\nu_0 }{2\pi}
                       \rho & 0\leq \rho\leq 2\pi \\
                        \mu_k+ \frac{\nu_k}{2\pi}(\rho-2\pi k) &~~ 2\pi k\leq \rho \leq 2\pi(k+1),\;\;\;\; k=1,....,P-1\\
                      \mu_P-  \frac{\mu_P}{2\pi}(\rho-2\pi P) & 2\pi P\leq \rho \leq 2\pi(P+1).
                                             \end{array}
\right.
\end{equation}
%
%
By imposing continuity the $\alpha_k$, $\mu_k$ integration constants are determined from $\beta_k$, $\nu_k$ as (see Appendix B),
\begin{equation}
\alpha_k=\sum_{j=0}^{k-1} \beta_j ,~~~\mu_k= \sum_{j=0}^{k-1}\nu_j.\label{definitionmupalphap}
\end{equation} 
These profiles for the $\hat{h}_4$ and $h_8$ functions were the ones taken in \cite{Lozano:2019zvg}-\cite{Lozano:2019ywa} for the construction of the 2d CFTs dual to the AdS$_3\times S^2\times \text{CY}_2$ solutions in \cite{Lozano:2019emq}. These results are summarised in Appendix \ref{AdS3}. The supergravity backgrounds can be trusted when $\beta_k,\nu_k,P$ are large. Taking $\beta_k,\nu_k$ to be  large controls the divergence of the Ricci scalar at the points where the sources are localised. In turn, these singularities are spread out for $P$ large.

The behaviour of the solutions defined by (\ref{profileh4sp}) and (\ref{profileh8sp}) (and $u=$ constant) at both ends of the $\rho$-interval is that of a superposition of D4 branes smeared on the $\text{CY}_2$ and D8 branes. Note that the same behaviour is obtained from a superposition of O4 and O8 orientifold fixed planes.  We will find an interesting interpretation for this behaviour in the following sections. 
Indeed, for very small values of $\rho$, the metric and dilaton read
\begin{equation}
\mathrm{d} s^2 \simeq  \rho^{-1} \left( \text{d}s^2_{\text{AdS}_2} + 4 \text{d}s^2_{S^3} \right) +  \text{d}s^2_{\text{CY}_2} +  \rho \, \mathrm{d} \rho^2 \, , \qquad e^{-2 \Phi} \simeq \rho^3 \, ,\label{forAppC1}
\end{equation}
while for $\rho \rightarrow 2 \pi (P+1)$ we have
\begin{equation}
\mathrm{d} s^2 \simeq x^{-1} \left( \text{d}s^2_{\text{AdS}_2} + 4 \text{d}s^2_{S^3} \right) + \text{d}s^2_{\text{CY}_2} + x\, \mathrm{d}x^2 \, , \\
e^{-2 \Phi} \simeq  x^3 \, ,\label{forAppC2}
\end{equation}
with $x=2\pi (P+1)-\rho$. We recognise these behaviours as those of a superposition of (smeared) D4 and D8 branes and/or O4 and O8 orientifold fixed planes.

Using the electric and magnetic charges discussed in the previous subsection to count, respectively, the colour and flavour brane charges, we find for $\rho$ in the $ [2\pi k, 2\pi (k+1)]$ interval,
\begin{eqnarray} 
&&Q_{\text{D0}}^{e\,(k)} =  \mu_k\, , \qquad  Q_{\text{D4}}^{e\,(k)} = \alpha_k \label{QD0D4}\\
&&Q_{\text{D4}'}^{m\,(k)} =  \beta_k\, , \qquad Q_{\text{D8}}^{m\,(k)}= \nu_k\, . \label{QD4'D8}
\end{eqnarray}
These equations show that  the constants $\alpha_k$, $\mu_k$, $\beta_k$, $\nu_k$ must be integer numbers.  
Moreover, they show that the D0 and D4 brane charges in each $ [2\pi k, 2\pi (k+1)]$ interval are equal to the total D8 and D4' brane charges 
in the $[0,2\pi k]$ previous intervals. Namely, $\alpha_k=\sum_{j=0}^{k-1} \beta_j $, $\mu_k= \sum_{j=0}^{k-1}\nu_j$.
We will find an interesting interpretation for this result when we discuss the dual field theory in section \ref{ADHM2}.

\subsection{Holographic central charge} \label{holocharge}

We compute the holographic central charge following the prescription in \cite{Lozano:2020txg}. In this section we consider general solutions with $u$ a linear function of $\rho$. We recall that as discussed in that reference the holographic central charge for an AdS$_2$ solution is to be interpreted as the number of vacuum states of the dual SCQM. The prescription in \cite{Lozano:2020txg} reads,
\begin{equation}
c_{\text{holo}}=\frac{3 V_{int}}{4\pi G_N}\, ,
\end{equation}
where $G_N=8\pi^6$ and $V_{int}$ is the volume of the internal space, which must be corrected by a dilaton term, following \cite{Macpherson:2014eza}. For the backgrounds described by (\ref{NS sector analytically continued}) and (\ref{RR sector analytically continued}), $V_{int}$ reads
\begin{equation}
V_{int}=\int \text{d}^8x\sqrt{e^{-4\Phi}\text{det} g_{8,ind}}=\frac{\text{Vol}_{\text{CY}_2} \text{Vol}_{S^3}}{4} \int_{0}^{2 \pi (P+1)} \mathrm{d} \rho\,  \Bigl(4\hat{h}_4 h_8-u'^2\Bigr)   \, ,
\end{equation}
from where, using that $\Upsilon \text{Vol}_{\text{CY}_2}=16\pi^4$, we find
\begin{equation}
\label{choloC}
c_{\text{holo}}=\frac{3}{4\pi}\int_0^{2\pi(P+1)}\text{d}\rho\, \Bigl(4h_4h_8-u'^2\Bigr)\, .
\end{equation}
We will refer to this expression in the next sections for our particular solutions with $u=$ constant.

\section{The dual superconformal quantum mechanics}\label{ADHM}

In this section we discuss the $\mathcal{N} = 4$ super-conformal quantum mechanical theories that we propose as duals to the AdS$_2$ solutions with the defining functions given by equations (\ref{profileh4sp}), (\ref{profileh8sp}) (and $u=$ constant). We provide a UV $\mathcal{N} = 4$  quantum mechanics, that conjecturally flows to a super conformal quantum mechanics dual to these backgrounds.

We start analysing a well-known particular solution in this class. This is the Abelian T-dual of the AdS$_3\times S^3\times$ CY$_2$ solution to Type IIB, along the $S^1$ fibre direction of the AdS$_3$ space. Thus, the dual SCQM to this solution arises as the IR fixed point of the quantum mechanical quiver that is obtained dimensionally reducing  the 2d QFT living in the D1-D5 system. 
Though many of the subtleties in our generic case do not appear in this simple setting, it is useful to study it first.
In fact, this example illustrates the field content that will appear in our quivers and sets up the discussion for the extension to more general ones. Appendix \ref{open-strings} contains a detailed account of the low-energy field content emerging from the brane web associated to our solutions. 

\subsection{Warm up: Quantum mechanics of the D0-D4 system}\label{ATD}

In order to motivate the construction of the quantum mechanics dual to our AdS$_2$ backgrounds we focus first on the Abelian T-dual of the D1-D5 system, whose near horizon geometry is the AdS$_3\times S^3\times$ CY$_2$ solution of Type IIB. In this case the T-duality is performed on the $S^1$ Hopf fibre contained in AdS$_3$, leaving an AdS$_2$ solution in our class where the $\rho$-direction lives in the T-dual circle. In this case $u$, $h_4$ and $h_8$ are the constant functions, 
\begin{equation}
u=16L^4M^2\, , \qquad h_4=4L^2M^4\, , \qquad h_8=4L^2\, ,
\end{equation}
and the solution reads
\begin{eqnarray}
&&\text{d}s^2=L^2\Bigl(\text{d}s^2_{\text{AdS}_2}+4\text{d}s^2_{S^3}\Bigr)+M^2 \text{d}s^2_{\text{CY}_2}+\frac{1}{4L^2}\text{d}\rho^2 \label{ATDmetric}\\
&& e^{-\Phi}=2L\\
&&H_{(3)}=-\frac{1}{2}\text{d}\rho \wedge \hat{\text{vol}}_{\text{AdS}_2}\\
&& F_{(4)}=-8L^2 \hat{\text{vol}}_{S^3}\wedge \text{d}\rho\\
&&F_{(8)}=8L^2 M^4 \hat{\text{vol}}_{S^3}\wedge \hat{\text{vol}}_{\text{CY}_2}\wedge \text{d}\rho\, ,
\label{ATDF8}
\end{eqnarray}
with $\rho\in [0,2\pi]$.
As in many other examples where T-duality is performed along a Hopf-fibre direction, the T-dual background preserves half of the supersymmetries of the original solution. In our case the SO$(4)_R$ symmetry of the AdS$_3\times S^3\times$ CY$_2$ solution is broken to SU$(2)_R$, the other SU$(2)$ becoming a global symmetry. Together with the SL$(2,\mathbb{R})$ isometry group of AdS$_2$, $\text{SL}(2,\mathbb{R})\times \text{SU}(2)$ span the bosonic subgroup of SU$(1,1|2)$. This is one of the two possible superconformal groups with 4 supercharges in one dimension, the other being D$(2,1;\alpha)$, $\alpha\neq -1,0$, for which the R-symmetry is two SU$(2)$'s, and thus corresponds to large superconformal symmetry.

The D1-D5 system gives rise upon T-duality to D0-D4 branes plus an extra F1-brane. The corresponding 1/4-BPS brane set-up is depicted in Table \ref{D0D4F1}.
\begin{table}[h]
\begin{center}
\begin{tabular}{c|c|c|c|c|c|c|c|c|c|c}
& $x^0$ & $x^1$ & $x^2$ & $x^3$ & $x^4$ & $x^5$ & $x^6$ & $x^7$ & $x^8$ & $x^9$\\
\hline
D0 & $-$ & & & & & $$ & & &$$ &$$  \\
\hline
 D4 &  $-$ &$-$ &$-$  &$-$  &$-$  & & $$ & $$ & $$ &  \\
\hline F1 &$-$ & $$ & $$ & $$ & $$ & $-$& $$ & & & $$ \\
\end{tabular}
\caption{Brane set-up associated to the D0-D4-F1 brane system T-dual to the D1-D5 system.} \label{D0D4F1}
\end{center}
\end{table}
The numbers of D0 and D4 branes are computed using equations (\ref{D0D4electric}). We obtain that $Q_{\text{D0}}^e=h_8$, $Q_{\text{D4}}^e=h_4$. There is also one F1-string that extends in the $\rho$-direction.
If before the T-duality the AdS$_3$ subspace is orbifolded by $\mathbb{Z}_N$, which is equivalent to introducing $N$ units of momentum in the D1-D5 system, breaking the supersymmetries to (0,4), then $N$ F1-strings are generated after the T-duality. These strings stretch in $\rho$ between $2\pi k$ and $2\pi (k+1)$, with $k=0,\dots, N-1$. 
The central charge of the D1-D5-wave system \cite{Kutasov:1998zh},
 \begin{equation}
c=6 \,Q_w Q_{\text{D1}} Q_{\text{D5}}
\end{equation}
is reproduced after the T-duality as
\begin{equation} \label{ccATD}
c= 6 \,Q_{\text{F}1} Q_{\text{D0}} Q_{\text{D4}}\, .
\end{equation}
As a consistency check for expression (\ref{choloC}) we can show that it reproduces this field theory result.



\begin{figure}[h!]
    \centering
    {{\includegraphics[width=1.15cm]{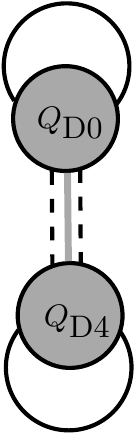} }}%
   \caption{Quiver quantum mechanics associated to the D0-D4-F1 brane system. The solid black lines represent (4,4) adjoint hypermultiplets, the grey line a (0,4) hypermultiplet and the dashed lines two $(0,2)$ Fermi multiplets. Circles represent ${\cal N}=(4,4)$ vector multiplets.}
\label{ATDquivers}
\end{figure}

We next discuss in some detail the quiver depicted in Figure \ref{ATDquivers}. 
This quiver  is the dimensional reduction of the 2d (4,4) quiver CFT that lives in the D1-D5 system. 
The usage of 2d (0,4) notation to describe 1d super quantum mechanics is widespread in the literature. The reader is referred to \cite{Hwang:2014uwa,Kim:2016qqs} for a detailed presentation of this description.
In our example this is further motivated by its explicit relation to the 2d (4,4) CFT living in the D1-D5 system. 
The detailed field content of the quiver depicted in Figure \ref{ATDquivers} is as follows (see Appendix \ref{open-strings}):
\begin{itemize}
\item Circles represent (4,4) vector multiplets. They are associated to gauge nodes. They come from open strings with both ends on the D0 or the D4 branes.
\item Black lines connecting one gauge node to itself represent (4,4) hypermultiplets in the adjoint representation of the gauge group. They also originate from open strings with both ends on the D0 or the D4 branes.
\item The grey line connecting the two gauge nodes represents a (0,4) bifundamental hypermultiplet. 
\item  The two dashed lines connecting  the two gauge nodes represent (0,2) bifundamental Fermi multiplets. These combine into a (0,4) Fermi multiplet. Together with the (0,4) bifundamental hypermultiplet they form a (4,4) hypermultiplet in the bifundamental representation of the two gauge groups. This originates from open strings stretched between the D0 and the D4 branes. The resulting quiver is thus (4,4) supersymmetric.
\end{itemize}



Having introduced our notation we now set out to describe the more general quiver quantum mechanics dual to the solutions 
defined by equations (\ref{profileh4sp}), (\ref{profileh8sp}). We will make use of the low-energy field content detailed in Appendix \ref{open-strings}.

\subsection{ADHM quantum mechanics with Wilson loops}\label{ADHM2}

The previous D0-D4-F1 brane system can be extended to include D4' and D8 branes while keeping the same number of supersymmetries, giving rise to the brane set-up depicted in Table \ref{Table brane web set type IIA3}.
We show in this and the next subsection that this brane set-up suggests an interpretation in terms of D0 and D4-brane self-dual instantons in the 5d $\mathcal{N}=1$ theory living in the D4'-D8 branes, with extra BPS Wilson loops. 

\begin{figure}[h!]
    \centering
    {{\includegraphics[width=5cm]{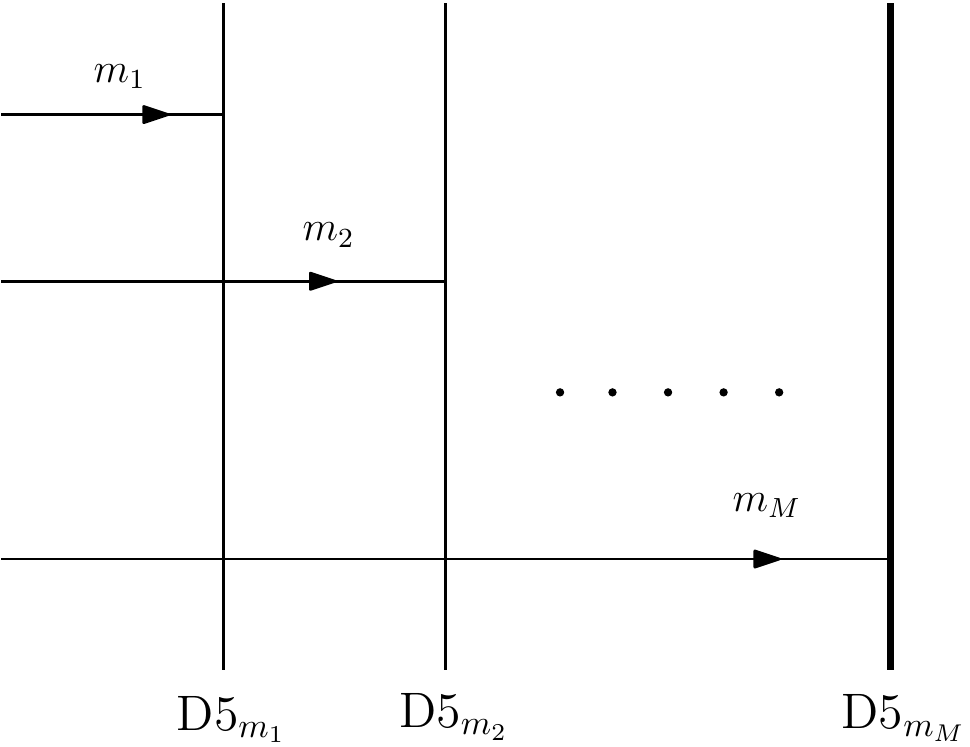} }}%
   \caption{Array of $M$ D5-branes with $(m_1, m_2, \dots m_M)$ F1-strings stretched between them and the $N$ D3-branes. Note that even if the branes are separated for illustration purposes they are actually coincident.}
\label{ArrayD3}
\end{figure}

\begin{figure}[h!]
    \centering
    {{\includegraphics[width=5cm]{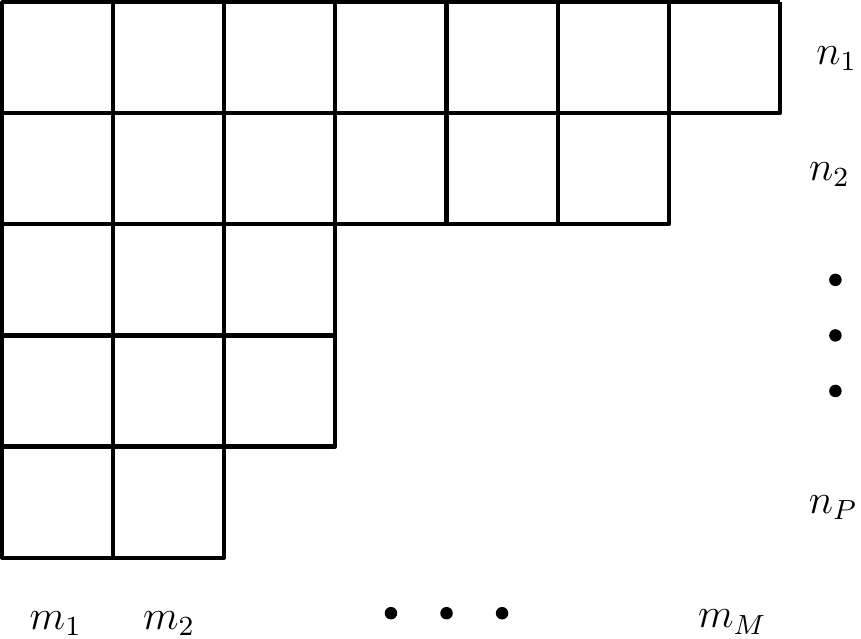} }}%
   \caption{Young tableau labelling the irreducible representations of U$(N)$.}
\label{youngtableau}
\end{figure}

We start recalling the brane realisation of Wilson loops in arbitrary U$(N)$ representations. This was worked out in \cite{Yamaguchi:2006tq,Gomis:2006sb,Gomis:2006im} for 4d $\mathcal{N}=4$ SYM. In \cite{Yamaguchi:2006tq,Gomis:2006sb} it was shown that a half-BPS Wilson loop in a U$(N)$ antisymmetric representation is described by an array of $M$ D5-branes with fundamental string charges dissolved in their worldvolumes. This is the realisation in the near horizon limit of a configuration of $M$ stacks of D5-branes separated by a distance $L$ from the $N$ D3-branes, with $(m_1,m_2,\dots m_M)$ F1-strings stretched between the stacks, as depicted in Figure \ref{ArrayD3}, in the limit $L\rightarrow \infty$.
The brane set-up is depicted in Table \ref{D3-D5-F1}. 
\begin{table}[h]
\begin{center}
\begin{tabular}{c|c|c|c|c|c|c|c|c|c|c}
& $x^0$ & $x^1$ & $x^2$ & $x^3$ & $x^4$ & $x^5$ & $x^6$ & $x^7$ & $x^8$ & $x^9$\\
\hline
D3 & $-$ & $-$ & $-$ & $-$ & & $$ & & &$$ &$$  \\
\hline
 D5 &  $-$ & &  &  &  & $-$ & $-$ & $-$ & $-$ & $-$ \\
\hline F1 &$-$ & $$ & $$ & $$ & $-$ & $$& $$ & & & $$ \\
\end{tabular}
\caption{Brane set-up associated to the D3-D5-F1 brane configuration that describes Wilson loops in the antisymmetric representation of U$(N)$ in 4d $\mathcal{N}=4$ SYM.} \label{D3-D5-F1}
\end{center}
\end{table}
In turn, in \cite{Gomis:2006sb,Gomis:2006im} it was shown that a half-BPS Wilson loop in a symmetric representation of U$(N)$ is described by an array of $P$ D3-branes with fundamental string charges dissolved in their worldvolumes. 
This is the realisation in the near horizon limit of a configuration of $P$ D3-branes, separated by a distance $L$ from a stack of $N$ coincident D3-branes, with $(n_1, n_2, \dots n_P)$ F1-strings stretched between the stacks, in the limit $L\rightarrow \infty$.
The F1-string charges dissolved in the different D5-branes (D3-branes) of the array $m_j$, $j=1,\dots, M$ ($n_i$, $i=1,\dots, P$), then realise a Wilson loop operator in the antisymmetric (symmetric) U$(N)$ representation labeled by the Young tableau depicted in Figure \ref{youngtableau}. 
This generalises the description of a Wilson loop in the fundamental representation in \cite{Rey:1998ik,Maldacena:1998im} to all other representations.

\begin{figure}[h!]
    \centering
    {{\includegraphics[width=5cm]{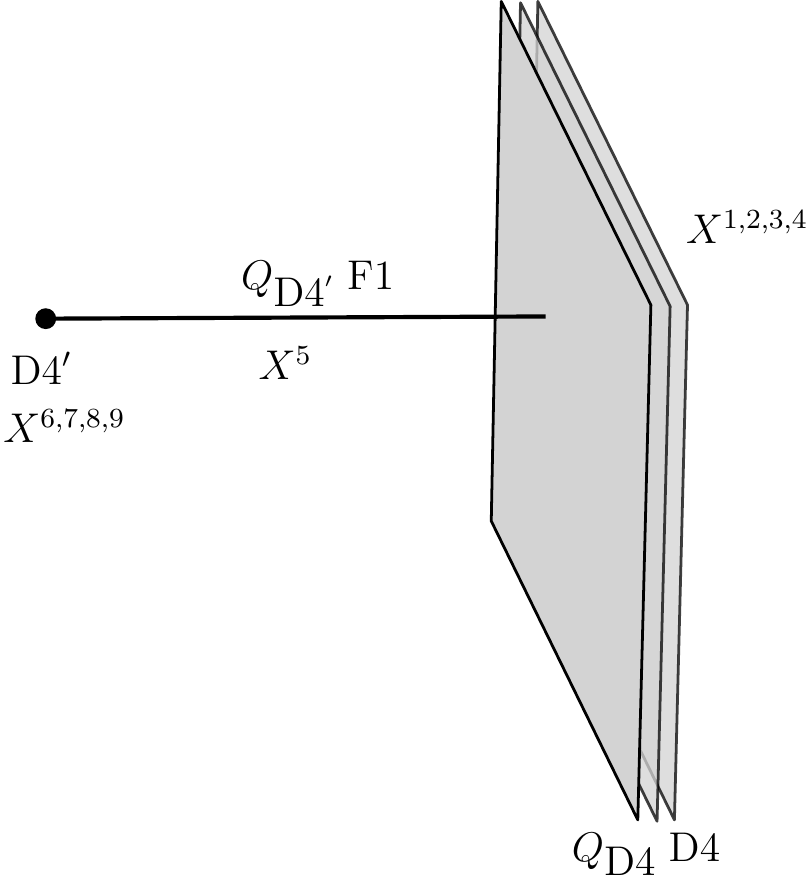} }}%
    \caption{Wilson loop in the $Q_{D4'}$-th antisymmetric representation of U$(Q_{\text{D4}})$.}
\label{D4primas}
\end{figure}

Let us now see how this is realised in our brane system. We start by considering the D4-D4'-F1 brane subsystem in Table \ref{Table brane web set type IIA3}. In this brane set-up the D4-D4'-F1 branes are distributed exactly as the D3-D5-F1 brane configuration that describes Wilson loops in antisymmetric representations in 4d $\mathcal{N}=4$ SYM, depicted in Table \ref{D3-D5-F1}. In other words,  the strings extending between the D4' and the D4 branes have as their lowest energy excitation a fermionic field. Integrating out this massive field leads to a Wilson loop in the antisymmetric representation.
Indeed, a BPS Wilson line can be introduced into the low-energy 5d SYM theory living on the D4-branes by probing the D4-branes with fundamental strings \cite{Rey:1998ik,Maldacena:1998im}. These can in turn be taken to originate on additional D4'-branes, orthogonal to the D4-branes. This can be described through the coupling
\begin{equation}
S_{\text{D4}}=T_4\int  \hat{F}_{(4)}\wedge A_t
\end{equation}
in the worldvolume effective action of the D4-branes. If the D4-branes are wrapped on the $\text{CY}_2$, as in our brane set-up, the D4'-branes must be orthogonal to them and must carry a magnetic charge 
\begin{equation}
Q_{\text{D4}'}=\frac{1}{(2\pi)^3}\int_{\text{CY}_2}\hat{F}_{(4)}\, .
\end{equation}
This charge is then equal to the number of F1-strings dissolved in the worldvolume of the D4-branes. For $Q_{\text{D4}'}$ D4'-branes this configuration, depicted in Figure \ref{D4primas},
describes then a Wilson loop in the $Q_{\text{D4}'}$-th antisymmetric representation of U$(Q_{\text{D4}})$, where $Q_{\text{D4}}$ is the number of D4-branes. BPS Wilson loops in arbitrary representations of U$(Q_{\text{D4}})$ can then be obtained adding arrays of D4'-branes with fundamental string charges dissolved in their worldvolumes, as in Figure \ref{ArrayD3}. 

Let us consider now the D0-D8-F1 brane subsystem in Table \ref{Table brane web set type IIA3}. In this brane set-up the D0-D8-F1 branes are again distributed exactly as the D3-D5-F1 and D4-D4'-F1 brane configurations described above. In this case a BPS Wilson line can be introduced into the low energy quantum mechanics living on the D0-branes by probing the D0-branes with fundamental strings, originating on D8-branes \cite{Chang:2016iji}. As discussed above, the fermionic string stretched between the D8 and the D0 branes implies that the Wilson loop will be in the antisymmetric representation. This is indeed what is inferred from the coupling
\begin{equation}
S_{\text{D0}}=T_0\int F_{(0)}\wedge A_t
\end{equation}
in the worldvolume effective action of the D0-branes. D8-branes with charge
\begin{equation}
Q_{\text{D8}}=2\pi F_{(0)}
\end{equation}
induce $Q_{\text{D8}}$ F1-strings dissolved in the worldline of the D0-branes. This describes then a Wilson loop in the $Q_{\text{D8}}$'th antisymmetric representation of U$(Q_{\text{D0}})$, for $Q_{\text{D0}}$ D0-branes. As for the D4-D4'-F1 subsystem described above, BPS Wilson loops in arbitrary representations of U$(Q_{\text{D0}})$ can then be obtained adding arrays of D8-branes with fundamental string charges dissolved in their worldvolumes, as depicted in Figure \ref{ArrayD3}.

Therefore, the complete D0-D4-D4'-D8-F1 brane system depicted in Table \ref{Table brane web set type IIA3} can be interpreted as describing Wilson loops in the $Q_{\text{D4}'}\times Q_{\text{D8}}$ antisymmetric representation of $\text{U}(Q_{\text{D4}})\times \text{U}(Q_{\text{D0}})$. 

The complete brane system has, however, a richer dynamics, as one must also consider the interactions between the D4-D4'-F1 and D0-D8-F1 subsystems. Indeed, the D0 and D4 branes can  be seen as instantons in the worldvolumes of the D4' and D8 branes, respectively. This is inferred from the couplings \cite{Douglas:1995bn}
\begin{equation}
S_{\text{D4}'}=T_4\int \text{Tr}[C_{(1)}\wedge F\wedge F]\, , \qquad S_{\text{D8}}=T_8\int \text{Tr}[C_{(5)}\wedge F\wedge F]\, ,
\end{equation}
in the D4' and D8 branes worldvolume effective actions. These show that a D0-brane can be absorbed by a D4'-brane and converted into an instanton, while a D4-brane wrapped on the $\text{CY}_2$ can be absorbed by a D8-brane and converted as well into an instanton. The one dimensional $\mathcal{N}=4$ gauged quantum mechanics living on the complete brane system would describe then the interactions between the two types of instantons and the two types of Wilson lines previously described. This generalises the ADHM quantum mechanics discussed in \cite{Tong:2014cha,Kim:2016qqs}. 

Indeed, in the previous references gauged quantum mechanics describing the interactions between D0-brane instantons and Wilson lines  in the 5d $\mathcal{N}=2$ SYM theory living in D4-branes were constructed. In our brane set-up we have extra D8-branes. These allow for extra D4-brane instantons wrapped on the $\text{CY}_2$. Moreover, the D8-branes introduce additional F1-strings ending on the D0-branes. Given this, we propose that the gauged quantum mechanics living on the complete D0-D4-D4'-D8-F1 brane set-up describes the interactions between instantons and Wilson lines in the 5d $\mathcal{N}=1$ SYM theory living in D4'-D8 branes.  

D4-D8 brane set-ups must include O8 orientifold fixed planes in order to flow to 5d fixed point theories in the UV \cite{Seiberg:1996bd,Brandhuber:1999np}. We have indeed seen in section \ref{set-up} that the behaviour of the solutions at both ends of the space is compatible with the presence of O8 orientifold fixed points. These could  provide for a fully consistent brane picture. 
In the next subsection, we construct quiver quantum mechanics that we propose describe instanton and Wilson line defects within the 5d Sp(N) D4-D8/O8 brane system of \cite{Seiberg:1996bd,Brandhuber:1999np}.

\subsection{The dual quiver quantum mechanics} 

In this section we propose quiver quantum mechanics supported by the D0-D4-D4'-D8-F1 brane system. The full dynamics 
is described in terms of the matter fields that enter in the description of the D0-D4-F1 system, introduced in section \ref{ATD}, plus additional fields that connect these branes with the D4' and the D8 branes. The extra fields are (twisted) (4,4) bifundamental hypermultiplets, coming from the open strings that connect the D4'-branes with the D0-branes and the D8-branes with the D4-branes, and (0,2) bifundamental Fermi multiplets, coming from the open strings that connect the D4'-branes with the D4-branes and the D8-branes with the D0-branes \cite{Tong:2014cha,Hwang:2014uwa}. This field content is detailed in Appendix \ref{open-strings}.

\begin{figure}[h!]
    \centering
    {{\includegraphics[width=11cm]{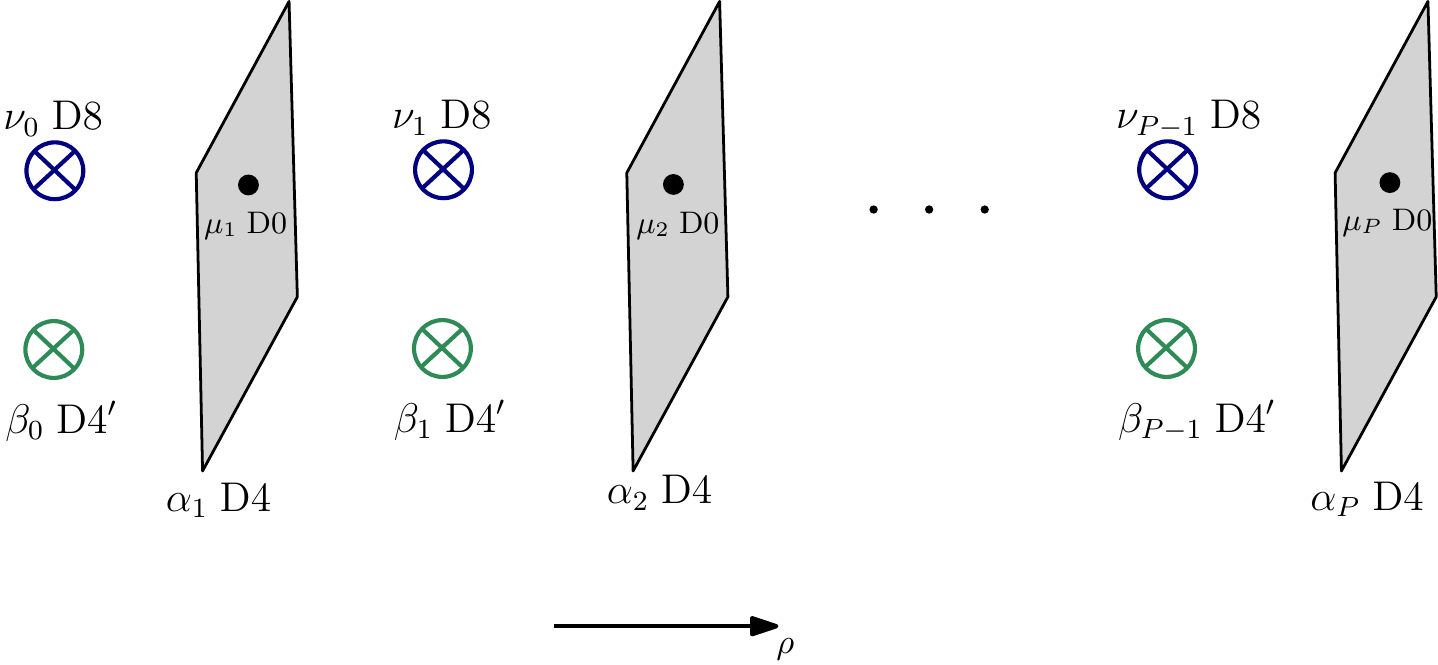} }}%
    \caption{Hanany-Witten like brane set-up associated to the quantised charges of the solutions.}
\label{one}
\end{figure}

Let us analyse in more detail the brane picture introduced in Table  \ref{Table brane web set type IIA3}.  
In subsection \ref{set-up} we found the following quantised charges
at each $[2\pi k, 2\pi (k+1)]$ $\rho$-interval,
\begin{eqnarray}
&&Q_{\text{D4}}^{e\,(k)}=\alpha_k=\sum_{j=0}^{k-1} \beta_j\, , \qquad Q_{\text{D0}}^{e\,(k)}=\mu_k=\sum_{j=0}^{k-1} \nu_j\, ,\\
&&Q_{\text{D4}'}^{m\,(k)}=\beta_k \, , \qquad Q_{\text{D8}}^{m\,(k)}=\nu_k \, , \qquad Q_{\text{F}1}^{e\, (k)}=1\, .
\end{eqnarray}
We can associate a Hanany-Witten like brane set-up to these charges, as depicted in Figure \ref{one}. 
In this figure there are 
$\mu_k$ D0-branes and $\alpha_k$ D4-branes, and orthogonal $\nu_{k}$ D8-branes and $\beta_{k}$ D4'-branes, in the $[2\pi k, 2\pi (k+1)]$, $k=1,\dots P$, $\rho$-intervals. As discussed in section  \ref{set-up} the D0 and D4 branes are interpreted as colour branes while the D4' and D8 branes play the role of flavour branes. In the last $[2\pi P,2\pi (P+1)]$ interval there are extra $-\alpha_P$, $-\mu_P$, D4' and D8 brane charges that cancel the total D4'  and D8 brane charges of the compact space. Note that these charges may originate from D4'/O4 and D8/O8 superpositions, consistently with the singularity structure at the end of the space. 

\begin{figure}[h!]
    \centering
    {{\includegraphics[width=11cm]{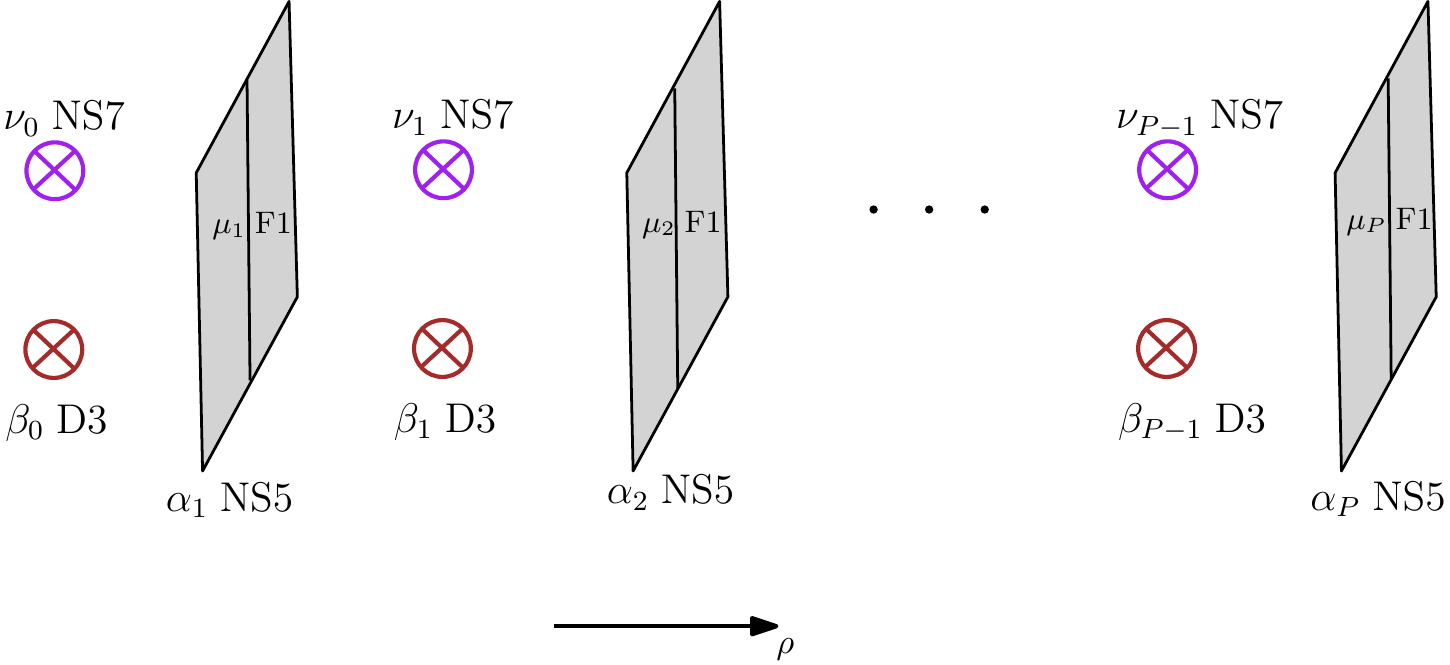} }}%
    \caption{Hanany-Witten brane set-up associated to the T+S duals of our solutions.}
\label{two}
\end{figure}

\begin{figure}[h!]
    \centering
    {{\includegraphics[width=11cm]{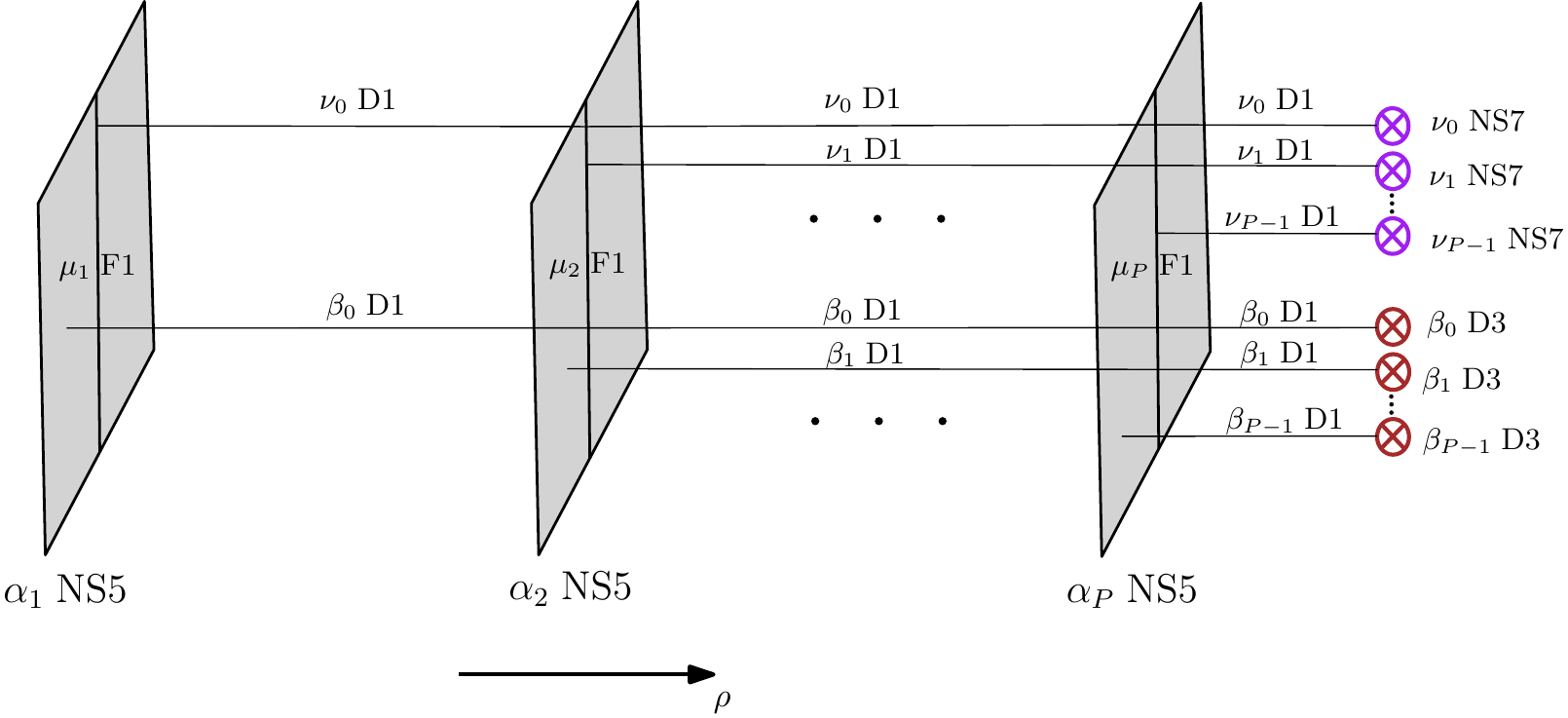} }}%
    \caption{Hanany-Witten brane set-up of Figure \ref{two} after Hanany-Witten moves.}
\label{three}
\end{figure}

The brane set-up depicted in Figure \ref{one} can be related to a Hanany-Witten brane set-up in Type IIB upon a T+S duality transformation. This is  depicted in Figure \ref{two}. 
The D0 and D4 branes are now F1 and NS5 branes, while the D8 and D4' branes become NS7 and D3 branes, respectively. In this brane set-up one can make Hanany-Witten moves that turn the configuration onto the one depicted in Figure \ref{three}. 
Here all the $\beta_k$ stacks of D3-branes and $\nu_k$ stacks of NS7-branes are now coincident, and there are $\alpha_k$ D1-branes and $\mu_k$ D1-branes ending on each stack of $\alpha_k$ NS5-branes and $\mu_k$ F1-strings, originating, respectively, from the $(\beta_0,\dots,\beta_{k-1})$ and $(\nu_0,\dots,\nu_{k-1})$ stacks of D3 and NS7-branes. 

\begin{figure}[h!]
    \centering
    {{\includegraphics[width=11cm]{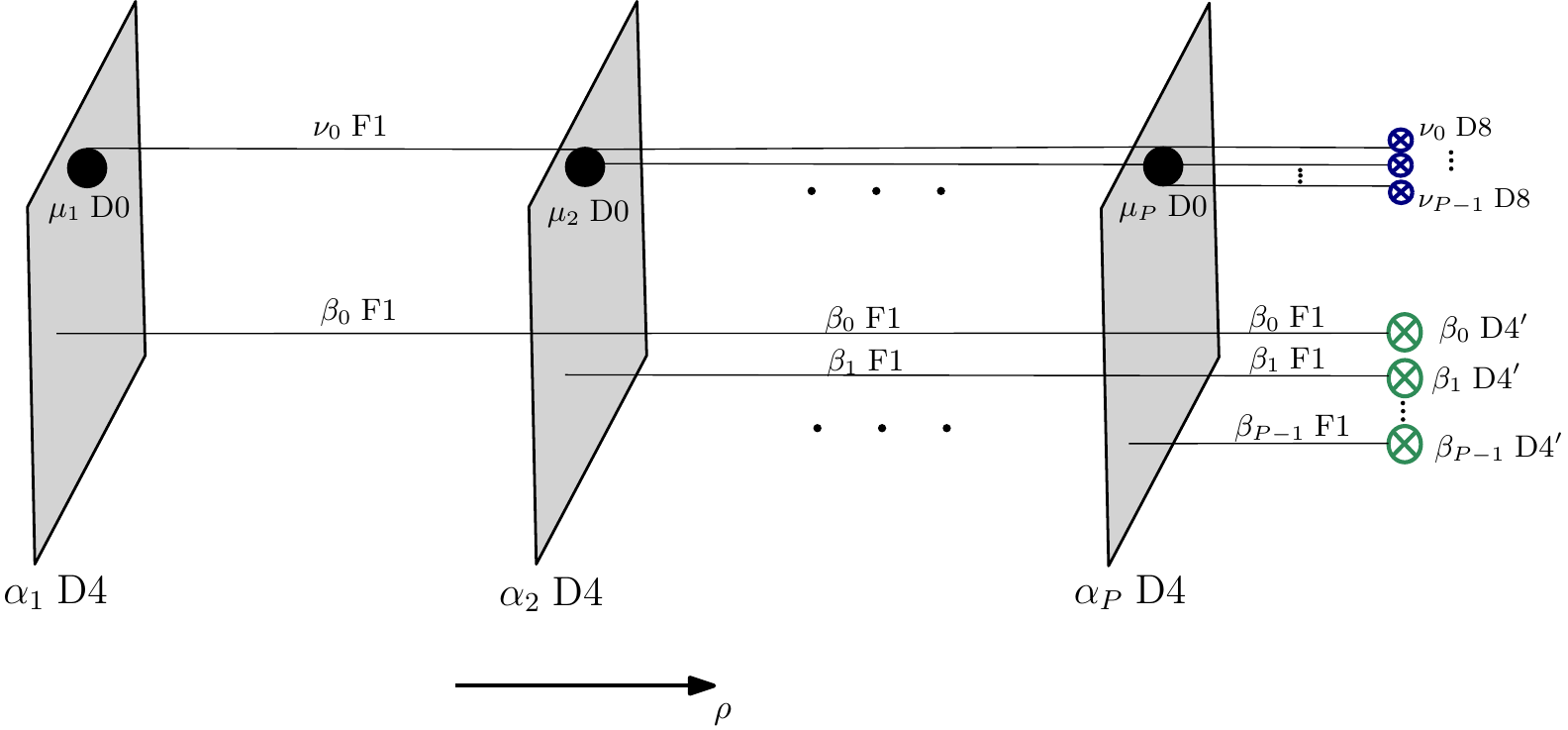} }}%
    \caption{Hanany-Witten like brane set-up equivalent to the brane configuration in Figure \ref{one}.}
\label{four}
\end{figure}

\begin{figure}[h!]
    \centering
    {{\includegraphics[width=10cm]{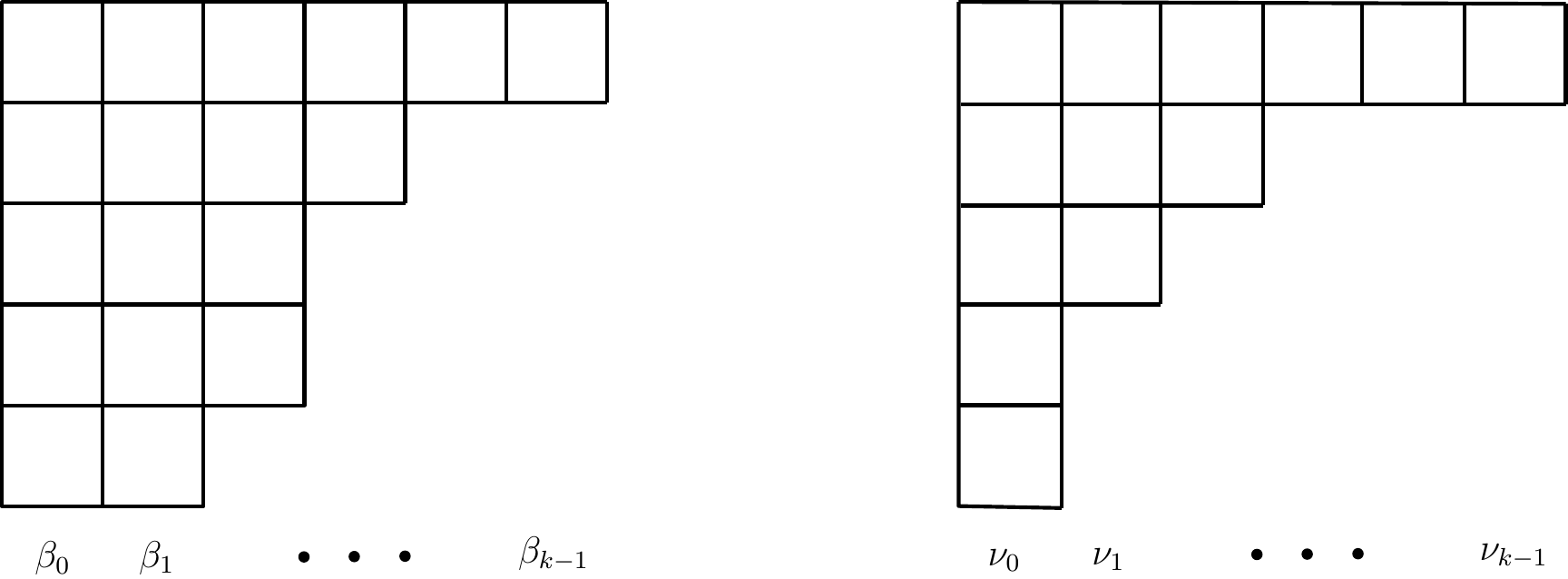} }}%
    \caption{Young tableaux labelling the irreducible representations of $U(\alpha_k)$ and $U(\mu_k)$.}
\label{youngtableauII}
\end{figure}

Back to Type IIA, the previous configuration is mapped onto the one depicted in Figure \ref{four}, where $\alpha_k$ F1-strings originating in $(\beta_0, \beta_1, \dots, \beta_{k-1})$ stacks of D4'-branes end on a given stack of $\alpha_k$ D4-branes, and  $\mu_k$ F1-strings originating in $(\nu_0, \nu_1, \dots, \nu_{k-1})$ stacks of D8-branes end on a given stack of $\mu_k$ D4-branes. 
This is exactly the description of U$(\alpha_k)$ and U$(\mu_k)$ Wilson loops in the antisymmetric representations $(\beta_0,\dots,\beta_{k-1})$ of U$(\alpha_k)$ and $(\nu_0,\dots,\nu_{k-1})$ of U$(\mu_k)$, that we discussed in the previous subsection.  The respective Young tableaux are depicted in Figure \ref{youngtableauII}. 

Therefore, our proposal is that the quantum mechanics dual to our AdS$_2$ solutions 
describes the interactions between Wilson loops in the $(\beta_0,\dots,\beta_{k-1})$ and $(\nu_0,\dots,\nu_{k-1})$ antisymmetric representations of the gauge groups $\text{U}(\alpha_k)\times \text{U}(\mu_k)$, and $\mu_k$ D0 and $\alpha_k$ D4 brane instantons, with $k=1,\dots,P$ \footnote{Note that if O4-O8 orientifold fixed planes are present at both ends of the $\rho$-interval  the gauge groups would  actually be $\text{Sp}(\alpha_1)$, $\text{Sp}(\mu_1)$ and $\text{Sp}(\alpha_P)$, $\text{Sp}(\mu_P)$ in the first and last $\rho$-intervals \cite{Bergman:2012kr}.}.

\begin{figure}[h!]
    \centering
    {{\includegraphics[width=10cm]{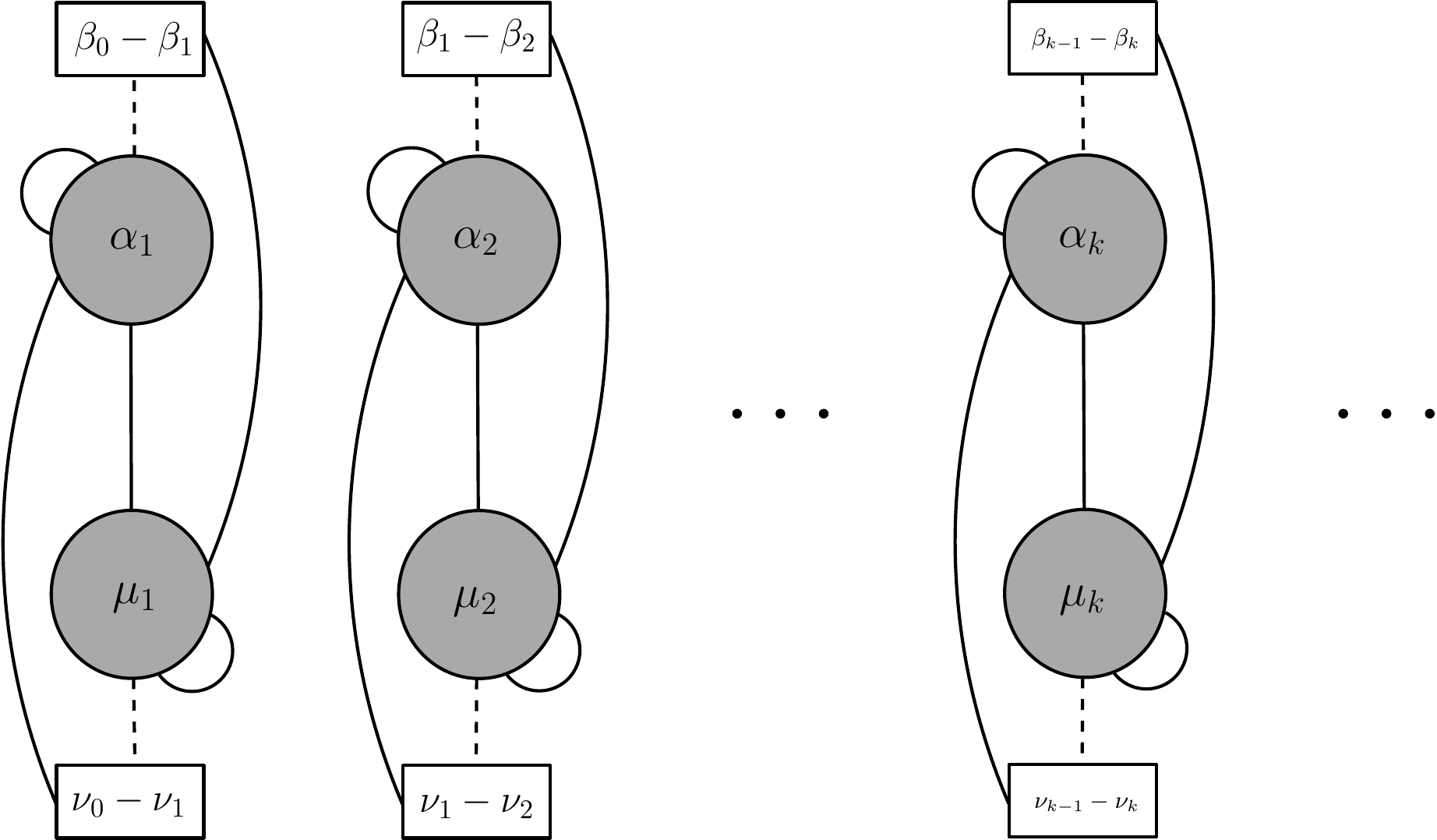} }}%
    \caption{Disconnected quivers describing the SCQMs dual to our solutions.}
\label{quivers}
\end{figure}

Our proposed quantum mechanics can be given a quiver-like description in terms of a set of disconnected quivers  showing the interactions between the D0-D4-D4'-D8 branes in each $\rho$-interval. These quivers can be read off directly from the brane set-up depicted in Figure \ref{one}, before the Hanany-Witten moves that connect the different intervals are made. Keeping in mind that D0 and D4 branes, and D4 and D8 branes, are connected by (4,4) hypermultiplets, and
D4' and D4 branes, and D8 and D0 branes, through (0,2) Fermi multiplets, all of them in the bifundamental representations of the respective groups, and that there are (4,4) vectors and (4,4) adjoint hypermultiplets at each gauge node (see Appendix \ref{open-strings}), we can depict the quivers shown in Figure \ref{quivers}. 
These quivers can be interpreted as partitions of the total number of D4' and D8 branes due to the insertions of the D0 and D4 brane instanton defects. This results in independent quantum mechanics living in the different D0-D4 brane instantons. In turn, the Wilson line defects do not show in the quivers, since the associated fermionic strings are massive. The integration of these Fermi fields leads to the insertion of the Wilson loop. 
Note that in these quivers we have taken into account that the number of D4' and D8 {\it source} branes in each interval  is given by the difference of the quantised charge in the given interval and that in the preceding one. This is implied by the source terms (\ref{D4 charges}), (\ref{D8 charges}), together with the derivatives
\begin{equation}
\label{bianchi-sources}
\hat{h}_4''=\frac{1}{2\pi}\sum_{k=1}^{P}(\beta_{k-1}-\beta_{k})\delta(\rho-2\pi k)\, , \qquad
h_8''=\frac{1}{2\pi}\sum_{k=1}^{P}(\nu_{k-1}-\nu_{k})\delta(\rho-2\pi k)\, ,
\end{equation}
derived from the $\hat{h}_4$ and $h_8$ functions defined by (\ref{profileh4sp}) and (\ref{profileh8sp}). Flavour groups in the last $\rho$-interval should be present associated to the D4' and D8 flavour branes at the end of the space.

Our suggested interpretation of the AdS$_2$ solutions in \cite{Lozano:2020bxo} as duals to line defect quantum mechanics living in the D4'-D8 brane system of \cite{Seiberg:1996bd,Brandhuber:1999np} is in agreement with the findings in \cite{Faedo:2020nol} (see also \cite{Dibitetto:2018gtk}). In these references it was shown that the solutions with $\text{CY}_2=\text{T}^4$ flow in the UV to the AdS$_6$ solution dual to the 5d CFT living in the D4'-D8 brane system. Accordingly, a defect interpretation in terms of D0-D4-F1 branes was given. Our analysis suggests that the D0 and D4 branes would find an interpretation in terms of instantons inside the D4' and D8 branes and that the F1-strings would arise in the near horizon limit from F1-strings stretched between the D0 and the D8 branes and the D4 and the D4' branes, realising Wilson loops in antisymmetric representations. Quiver quantum mechanics describing line defects in 5d gauge theories realised in $(p,q)$ 5-brane webs have been proposed in \cite{Assel:2018rcw,Assel:2019iae} (see also \cite{Hwang:2014uwa,Chang:2016iji}). It would be interesting to clarify the relation between these SCQM and the ones proposed in this paper.

\subsection{Quantum mechanical central charge}

One possible check of our proposed quivers would require that a notion of central charge existed for the superconformal quantum mechanics, that could be matched with the holographic central charge constructed in section  
\ref{holocharge}. Such a notion indeed exists for 1d quiver CFTs that originate from 2d $\mathcal{N}=(0,4)$ CFTs upon dimensional reduction \cite{Lozano:2020txg}. In that case one can measure the number of vacua of the 1d CFT using the same expression that defines the central charge of the 2d CFT, in terms of the two-point U$(1)_R$ current correlation function (see for example \cite{Putrov:2015jpa}), 
\begin{equation} \label{2dcentralcharge}
c=6(n_{hyp}-n_{vec})\, ,
\end{equation}
where $n_{hyp}$ is the number of $\mathcal{N}=(0,4)$ hypermultiplets and $n_{vec}$ the number of $\mathcal{N}=(0,4)$ vector multiplets in the UV description (of either the 1d or the 2d CFT). It was shown in \cite{Lozano:2020txg}  that for the class of AdS$_2$ solutions considered therein the result matches the holographic calculation for long quivers with large ranks. We show next that equation (\ref{2dcentralcharge}) matches as well the holographic result for the quiver quantum mechanics defined in the previous subsection, not originating from 2d CFTs. For these quivers  $n_{hyp}$ and $n_{vec}$ are the numbers of $\mathcal{N}=(0,4)$ hypermultiplets and $\mathcal{N}=(0,4)$ vector multiplets of the quantum mechanics. 

As a first check we show that equation (\ref{2dcentralcharge}) reproduces the right central charge of the D0-D4-F1  system, T-dual to the D1-D5-wave system. In this case, with $N$-waves,
\begin{eqnarray}
&&n_{hyp}=N Q_{\text{D0}}Q_{\text{D4}}+N (Q_{\text{D0}}^2+Q_{\text{D4}}^2)\, , \qquad n_{vec}=N(Q_{\text{D0}}^2+Q_{\text{D4}}^2)\nonumber\\
&&c=6(n_{hyp}-n_{vec})=6NQ_{\text{D0}}Q_{\text{D4}}\, .
\end{eqnarray}
This is in exact agreement with equation (\ref{ccATD}) and (\ref{choloC}).

\begin{figure}[h!]
    \centering
    {{\includegraphics[width=9cm]{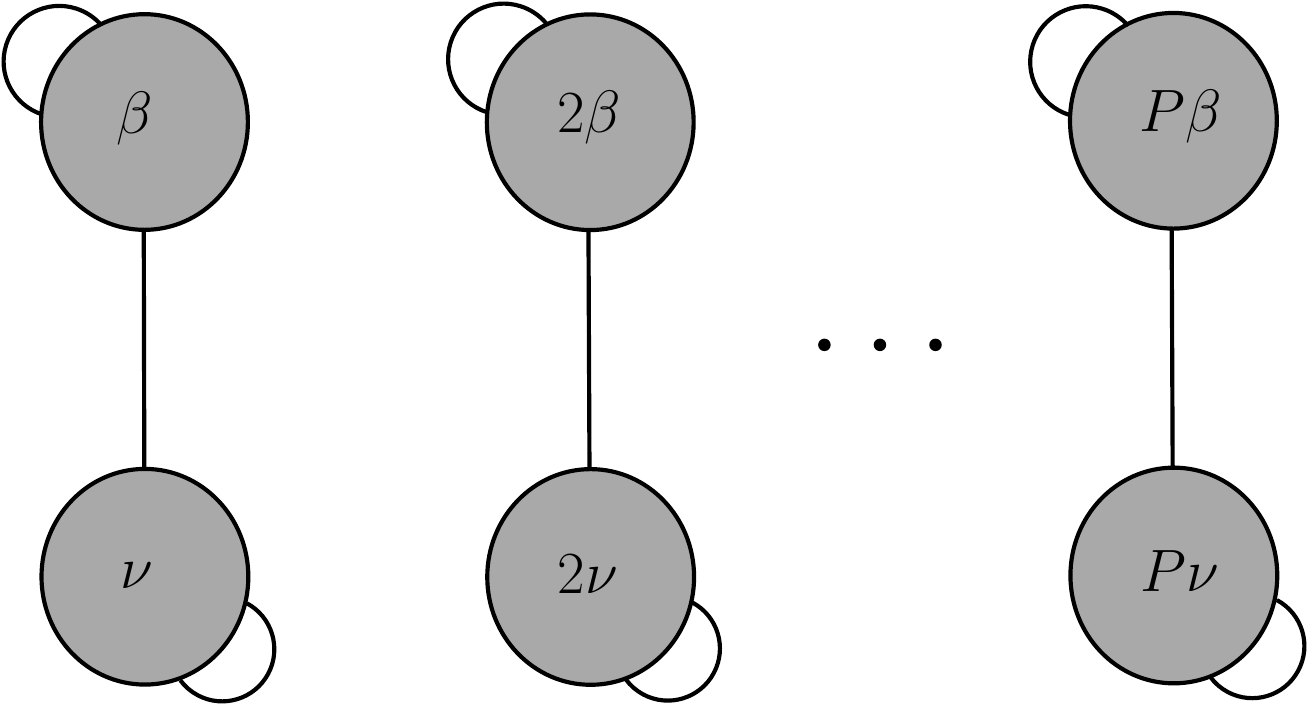} }}%
  \caption{Quiver mechanics associated to the backgrounds obtained from eqs.(\ref{profileh4exampleII})-(\ref{profileh8exampleII}).}
\label{Example}
\end{figure}

As a second example we consider the following profiles for the $\hat{h}_4$ and $h_8$ linear functions:
 \begin{equation} \label{profileh4exampleII}
\hat{h}_4(\rho)=\Upsilon h_4(\rho)
                    =\Upsilon\left\{ \begin{array}{ccrcl}
                       \frac{\beta }{2\pi}
                       \rho & 0\leq \rho\leq 2\pi P\\
                      \frac{\beta P}{2\pi}(2\pi (P+1) -\rho)\,, & 2\pi P\leq \rho \leq 2\pi(P+1).
                                             \end{array}
\right.
\end{equation}
\begin{equation} \label{profileh8exampleII}
h_8(\rho)
                    =\left\{ \begin{array}{ccrcl}
                       \frac{\nu }{2\pi}
                       \rho & 0\leq \rho\leq 2\pi P \\
                      \frac{\nu P}{2\pi}(2\pi( P+1) -\rho)\, , & 2\pi P\leq \rho \leq 2\pi(P+1).
                                             \end{array}
\right.
\end{equation}
The corresponding $\mathcal{N}=4$ quantum mechanical quiver is the one depicted in Figure \ref{Example}\footnote{Flavour groups coupled to the last gauge nodes should be present associated to the D4' and D8 flavour branes at the end of the space. Their contribution to the central charge is subleading and has not been incorporated into the (therefore approximate) expressions for the number of hypermultiplets and vector multiplets.}.

\noindent Substituting in (\ref{2dcentralcharge}) we find
\begin{eqnarray}
& & n_{hyp}\sim \sum_{j=1}^P j^2\Bigl(\beta^2 +\nu^2 +\beta\nu\Bigr),\;\;\;\; n_{vec}\sim\sum_{j=1}^P j^2 (\beta^2+\nu^2),\label{centralex}\nonumber\\
& & c\sim  \beta\nu P(P+1)(2P+1)  \sim 2\beta\nu P^3 \, , \qquad  c_{hol,1d}=2\beta\nu P^2 (P + 1)\sim 2\beta\nu P^3.\nonumber
\end{eqnarray}
We then see that in the holographic limit (large $P$, $\nu$, $\beta$) the field theory and holographic central charges  coincide. One can elaborate many other examples in which the two calculations are also shown to agree. 

 In order to justify this agreement one can draw a comparison between the central charge calculation proposed in (\ref{2dcentralcharge}) and the dimension of the Higgs branch of the quantum mechanics calculated in \cite{Denef:2002ru,Ohta:2014ria,Cordova:2014oxa}. In these references the following formula is used to compute the dimension of the Higgs branch for $\mathcal{N}=4$ quantum mechanics with gauge group $\Pi_v \text{U}(N_v)$ 
\begin{equation}
\mathcal{M}=\sum_{v,w}N_v N_w-\sum_v N_v^2+1\, .
\end{equation}
In this formula $N_w$ stands for the ranks of the colour groups adjacent to a given colour group of rank $N_v$. Our expression in  
(\ref{2dcentralcharge}) extends this fomula for counting the degrees of freedom of the quantum mechanics to more general quivers including flavours. Moreover, our quivers need not be related by dimensional reduction to 2d CFTs, as the quivers discussed in \cite{Lozano:2020txg}, where this formula was shown to reproduce the central charge of 1d CFTs obtained from 2d CFTs upon dimensional reduction.

In the next section we further elaborate on the holographic central charge discussed in section \ref{holocharge}, and relate it to an extremisation principle.

\section{Holographic central charge, electric-magnetic charges  {and a minimisation principle}} \label{holomini}

In this section we show that the holographic central charge of the AdS$_2$ solutions discussed in this paper can be related to the product of the RR electric and magnetic charges. {The second result of this section is to show that it can also be obtained through a minimisation principle}. The first relation was already encountered for the $\mathcal{N}=4$ AdS$_2$ solutions studied in  \cite{Lozano:2020txg}, and, as in that case, it generalises an argument for AdS$_2$ gravity coupled to a gauge field put forward in \cite{Hartman:2008dq}.{ Our second result is a minimisation principle that allows to obtain the holographic central charge in the spirit of \cite{Couzens:2018wnk,Hosseini:2019ddy}, by minimising a functional defined as the integral of various geometrical forms. In analogy with the findings in  \cite{Lozano:2020txg}, we show that these geometric forms can be directly related to the RR electric and magnetic fluxes of the background. In contrast with  \cite{Lozano:2020txg}, we use the Maxwell (instead of Page) fluxes to establish the connection.}

\subsection{Relation with the electric-magnetic charges}\label{xxuu}

As it happened for the AdS$_2$ solutions in  \cite{Lozano:2020txg}, there exists an interesting relation between the holographic central charge of our AdS$_2$ solutions, given by equation \eqref{choloC}, and the electric and magnetic charges of the underlying Dp-branes. Using our definitions (\ref{electric-charges}) and (\ref{magnetic-charges}) for the electric and magnetic charges of Dp-branes and taking the absolute value of the charges, we find that the quantity 
\begin{equation}
\mathcal{Q}=\sum_p Q_{\text{Dp}}^e Q_{\text{Dp}}^m
\end{equation}
is proportional to the holographic central charge, up to a boundary term. More concretely, we find
\begin{eqnarray}
\label{electric-magnetic}
\mathcal{Q}&=&\frac{\text{Vol}_{\text{CY}_2}\text{Vol}_{S^3} \text{Vol}_{\text{AdS}_2}}{(2\pi)^8}  \int \text{d} \rho \Bigl[
2 \hat{h}_4 h_8 - \frac{(u')^2}{2} 
-\frac{(\rho-2\pi k)}{2}(4\hat{h}_4h_8-(u')^2)\left(\frac{\hat{h}_4'}{\hat{h}_4}+\frac{h_8'}{h_8}\right)\nonumber\\
& &+\frac{u^2-2(\rho-2\pi k)uu'}{4}\left(\frac{\hat{h}_4'^2}{\hat{h}_4^2}+\frac{h_8'^2}{h_8^2}\right)\Big]
\end{eqnarray}
where the sum has been taken over the D0-D4-D4'-D8 branes of our backgrounds. This expression needs to be regularised in order to obtain a finite result, given the integral over the infinite AdS$_2$ space present in the computation of the electric charges. We  can regularise it for instance as in (\ref{regpres}). 
\\
Let us now use the BPS equation $u''=0$ on the expression (\ref{electric-magnetic}).  We may think about the application of the BPS equation as an extremisation of the quantity $\mathcal{Q}$. We find
\begin{eqnarray}
\label{electric-magnetic2}
\mathcal{Q}&=&\frac{\text{Vol}_{\text{CY}_2}\text{Vol}_{S^3} \text{Vol}_{\text{AdS}_2}}{(2\pi)^8}  \int \text{d} \rho \Bigl[
4 \hat{h}_4 h_8 -( u')^2 
+\frac{u^2 - 2 (\rho-2\pi k) u u'}{4}\left(\frac{\hat{h}_4''}{\hat{h}_4}+\frac{h_8''}{h_8}\right)\nonumber\\
& &-\partial_{\rho} \Bigl( 2 (\rho-2\pi k) \hat{h}_4 h_8 - \frac{u u'}{2 } + \frac{u^2 - 2 (\rho-2\pi k) u u'}{4}\left( \frac{(\hat{h}_4 h_8)'}{\hat{h}_4 h_8}\right) \Bigr) \Bigr].
%
\end{eqnarray}
Finally, we use the expressions (\ref{bianchi-sources}) to obtain,
\begin{eqnarray}
\label{electric-magnetic3}
\mathcal{Q}&=&\frac{\text{Vol}_{\text{CY}_2}\text{Vol}_{S^3} \text{Vol}_{\text{AdS}_2}}{(2\pi)^8}  \int \text{d} \rho \Bigl[
4 \hat{h}_4 h_8 -( u')^2 +\\
& & \frac{u^2 - 2 (\rho-2\pi k) u u'}{8\pi}\sum_{k=1}^{P}\left(\frac{(\beta_{k-1}-\beta_{k})}{{h}_4}+\frac{(\nu_{k-1}-\nu_{k})}{h_8}\right)\delta(\rho-2\pi k)\nonumber\\
& & 
-\partial_{\rho} \Bigl( 2 (\rho-2\pi k) \hat{h}_4 h_8 - \frac{u u'}{2 } + \frac{u^2 - 2 (\rho-2\pi k) u u'}{4}\left( \frac{(\hat{h}_4 h_8)'}{\hat{h}_4 h_8} \right)\Bigr) \Bigr]\nonumber.
%
%
%
%
%
%
\end{eqnarray}
In the absence of sources, the second line in (\ref{electric-magnetic3}) vanishes and the quantity $\mathcal{Q}$ in (\ref{electric-magnetic}) coincides, up to a finite boundary term with the holographic central charge in (\ref{choloC}). 

Consider now solutions with $u'=0$ as in the rest of this paper. In the presence of sources, the second line in eq.(\ref{electric-magnetic3}) is,
\begin{equation}
\frac{u_0^2}{8\pi}\sum_{k=1}^{P}\left(\frac{(\beta_{k-1}-\beta_{k})}{\alpha_k}+\frac{(\nu_{k-1}-\nu_{k})}{\mu_k}\right).\nonumber
\end{equation}
This is subleading in the regime of large parameters, with respect to the term in the first line of (\ref{electric-magnetic3}). Interestingly, the boundary term gives a divergent contribution similar to that found in  \cite{Lozano:2020txg}.
In fact, for  $\hat{h}_4$ and $h_8$ given by equations (\ref{profileh4sp}) and (\ref{profileh8sp}), we find
\begin{eqnarray}
& & \int_{0}^{2\pi (P+1)}\partial_\rho \Bigl(2(\rho-2\pi k) \hat{h}_4 h_8 + u^2 \frac{(\hat{h}_4h_8)'}{4\hat{h}_4h_8}\Bigr) =-\lim_{\epsilon\rightarrow 0}\frac{u_0^2}{8\pi \epsilon}(\alpha_P +\beta_0+\mu_P+\nu_0)+ O(\epsilon^2)
\nonumber\\
&= & \lim_{\epsilon\rightarrow 0}\frac{u_0^2}{8\pi \epsilon}(Q_{\text{D4}'}^{\text{sources}}+Q_{\text{D8}}^{\text{sources}}),
\end{eqnarray}
where $\hat{h_4}(\rho=0)=h_8(\rho=0)=\epsilon$, and the same at $\rho=2\pi (P+1)$, with $\epsilon\rightarrow 0$. The boundary term thus yields a divergence directly related to the presence of the D4' and D8 brane {\it sources} of the background.
%
%
%
%
%
In summary, once regularised,  equation (\ref{electric-magnetic}) is proportional, up to a boundary term, to the holographic central charge in (\ref{choloC}). 

As in \cite{Lozano:2020txg}, we find an explicit realisation of the proposal in \cite{Hartman:2008dq}, which relates the central charge in the algebra of symmetry generators of AdS$_2$ with an electric field to the square of the electric field. This proposal is realised in our fully consistent string theory set-up, where various branes with electric and magnetic charges enter the calculation.
 We show next that we can also derive the central charge by extremising an action functional constructed out of the RR Maxwell fluxes.
\subsection{An action functional for the central charge}
We showed in \cite{Lozano:2020txg} that a natural way to define an action functional from which the central charge can be derived through an extremisation principle is to put ``off-shell'' the electric and magnetic RR Page fluxes of the associated background. We implement next this idea in our current backgrounds following closely the derivation in \cite{Lozano:2020txg}, with the  difference that in the present case we use the Maxwell fluxes.

In fact,  we use the fluxes  $F_0, F_2, F_4$ defined in (\ref{RR sector analytically continued})--as in the rest of the paper we consider the case $H_2=0$ and $\hat{h}_4(\rho)$. We also use the dual fluxes $F_6, F_8, F_{10}$ defined  as Hodge duals below (\ref{RR sector analytically continued}).

On these fluxes we impose the ``restriction'' procedure explained in \cite{Lozano:2020txg}. Basically, we define forms constructed from the Maxwell fluxes, after excising from them the AdS$_2$ part. In this way we define the forms
$[{\cal J}_0, \tilde{{\cal J}}_0, {\cal J}_4, {\cal \tilde{J}}_4, {\cal F}_4, {\cal \tilde{F}}_4, {\cal F}_8, \tilde{{\cal F}}_8 ]$. To be concrete, the Maxwell fluxes of our backgrounds and the ``restricted'' forms read,
\begin{equation}
 \begin{split}
F_{(0)}={\mathcal J}_0, \qquad \qquad
F_{(2)}=-\mathcal{\tilde{J}}_0\;\hat{\text{vol}}_{\text{AdS}_2}&,\qquad\qquad
F_{(4)}={\mathcal J}_{4}-\mathcal{\tilde{J}}_{4},\\
F_{(6)}={\mathcal F}_{4}\wedge\hat{\text{vol}}_{\text{AdS}_2}+\mathcal{\tilde{F}}_{4}\wedge\hat{\text{vol}}_{\text{AdS}_2},
\qquad\qquad
F_{(8)}={\mathcal F}_{8},&\qquad\qquad
F_{(10)}=\tilde{{\mathcal F}}_{8}\wedge\hat{\text{vol}}_{\text{AdS}_2},
 \end{split}
\end{equation}

\begin{equation}\label{forms}
\begin{split}
{\mathcal J}_0 &= h'_8 \, , \qquad\qquad
\mathcal{\tilde{J}}_0= \frac{1}{2} \bigg(h_8+\frac{u'uh_8'}{4\hat{h}_4h_8-u'^2} \bigg) ,  \qquad\qquad\mathcal{J}_4 = - \hat{h}_4'\;\hat{\text{vol}}_{\text{CY$_2$}} ,\\\mathcal{\tilde{J}}_4&=\left(2h_8+\frac{uu'\hat{h}_4'-\hat{h}_4u'^2}{2\hat{h}_4^2}\right)\hat{\text{vol}}_{S^3}\wedge\text{d}\rho,\qquad\qquad\mathcal{F}_4= \frac{h_8\hat{h}_4'u^2}{\hat{h}_4(4\hat{h}_4h_8-u'^2)} \hat{\text{vol}}_{S^3}\wedge\text{d}\rho,
 \\ \mathcal{\tilde{F}}_4&=\frac{1}{2} \bigg(\hat{h}_4+\frac{u'u\hat{h}_4'}{4\hat{h}_4h_8-u'^2} \bigg)
\;\hat{\text{vol}}_{\text{CY$_2$}},
\qquad
{\mathcal F}_{8} =\left(2\hat{h}_4+\frac{uu'h_8'-h_8u'^2}{2h_8^2}\right)\hat{\text{vol}}_{\text{CY$_2$}}\wedge\hat{\text{vol}}_{S^3} \wedge \mathrm{d} \rho, \\{\mathcal{\tilde{F}}}_{8} &=-\frac{h_8'\hat{h}_4u^2}{h_8(4\hat{h}_4h_8-u'^2)}   \hat{\text{vol}}_{\text{CY$_2$}}\wedge \hat{\text{vol}}_{S^3}\wedge\text{d}\rho.
\end{split}
\end{equation}
With the forms as in \eqref{forms}, we define the functional,
\begin{equation}\label{functionalcC}
\begin{split}
{\cal C}
&=\frac{1}{2}\int_{X_8}[\mathcal{\tilde{J}}_4\wedge \mathcal{\tilde{F}}_4+\mathcal{J}_0\wedge \tilde{\mathcal{F}}_8+\mathcal{F}_4\wedge \mathcal{J}_4+\mathcal{\tilde{J}}_0\wedge \mathcal{F}_8)]\\
&=\int_{X_8}  \left(\hat{h}_4 h_8 - \frac{1}{8}\left[u^2(\frac{\hat{h}_4'^2}{\hat{h}_4^2}+\frac{h_8'^2}{h_8^2})- 2u u' (\frac{\hat{h}_4'}{\hat{h}_4} +\frac{h_8'}{h_8}) + 2 u'^2  \right]\right){\hat{\text{vol}}}_{\text{CY}_2} \wedge  {\hat{\text{vol}}}_{S^3} \wedge \text{d}\rho.	
\end{split}
\end{equation}
We now extremise this functional by imposing the Euler-Lagrange equation for  $u(\rho)$,
\begin{eqnarray}
	2u''=u\left(\frac{\hat{h}_4''}{\hat{h}_4}+\frac{h_8''}{h_8}\right).
\end{eqnarray}
In  the absence of sources, this equation implies
 \begin{equation}
 \begin{split}
 	h_8''=0,\qquad \hat{h}_4''=0,\qquad u''=0.
 \end{split}
 \end{equation}
 In this case the functional ${\cal C}$ is proportional to the central charge in (\ref{choloC}). The functional in (\ref{functionalcC}) can be written as
 \begin{eqnarray}
 & & 4 {\cal C}= \int_{X_8} \left(4 \hat{h}_4 h_8 -(u')^2 -\frac{ u^2 }{2}\left(\frac{\hat{h}_4''}{\hat{h}_4}+\frac{h_8''}{h_8} \right) +\partial_\rho \left[\frac{u^2}{2}  \left(\frac{\hat{h}_4'}{\hat{h}_4}+\frac{h_8'}{h_8} \right) \right]\right){\hat{\text{vol}}}_{\text{CY}_2} \wedge  {\hat{\text{vol}}}_{S^3} \wedge \text{d}\rho .\label{mmll}\nonumber
 \end{eqnarray}
Using the expressions in (\ref{bianchi-sources}) and following the procedure described  below (\ref{electric-magnetic2}), we find for constant $u=u_0$
\begin{eqnarray}
  4 {\cal C}&=& \int_{X_8}\left( 4 \hat{h}_4 h_8 -(u')^2  -\frac{u_0^2}{4\pi} \sum_{k=1}^{P} \left( \frac{\beta_{k-1}-\beta_k}{h_4} +\frac{\nu_{k-1}-\nu_k}{h_8}\right)\delta(\rho-2\pi k)\right) {\hat{\text{vol}}}_{\text{CY}_2} \wedge  {\hat{\text{vol}}}_{S^3} \wedge \text{d}\rho\nonumber\\
& & - \lim_{\epsilon\to 0} \frac{u_0^2}{2\epsilon}(\mu_P+\nu_0 +\alpha_P+\beta_0) {\text{Vol}}_{\text{CY}_2}  {\text{Vol}}_{S^3}.
\end{eqnarray}
As in section \ref{xxuu}, we find that up-to a boundary term and the subleading source-term, the extremisation of the functional  ${\cal C}$ in (\ref{functionalcC}) is proportional to the holographic central charge (\ref{choloC}).
The functional ${\cal C}$ is defined in terms of the Maxwell fields of the background. This development extends the ideas of \cite{Couzens:2018wnk,Hosseini:2019ddy}  to the present case, for manifolds with boundary in  the presence of sources.

\section{Conclusions}

In this paper we have studied the AdS$_2\times S^3\times \text{CY}_2\times I_{\rho}$ solutions to massive Type IIA supergravity recently constructed in \cite{Lozano:2020bxo} and proposed SCQM dual to them. Our solutions can be thought of as describing the background near a string like defect inside AdS$_6$. Conversely, they can be thought of as dual to the Quantum Mechanics describing the excitations of $(0+1)$-dimensional defects in a five dimensional SCFT.

We have started by identifying the 1/8-BPS brane set-up that underlies the solutions. This is the D0-D4-D4'-D8-F1 brane intersection studied in \cite{Dibitetto:2018gtk,Faedo:2020nol}.  From the study of this brane set-up we have  revealed that the D0 and D4 branes have an interpretation as instantons in the worldvolumes of the D4' and D8 branes. Accordingly, these branes are counted by their electric charges. In turn, the F1-strings provide for Wilson lines in the antisymmetric representations of the D0 and D4 branes gauge groups. This generalises the constructions in  \cite{Yamaguchi:2006tq,Gomis:2006sb} for 4d $\mathcal{N}=4$ SYM to our more complicated brane configurations. 

We have 
proposed explicit quiver quantum mechanics dual to our solutions, that generalise previous constructions dual to AdS$_2$ solutions in 
\cite{Lozano:2020txg}. In \cite{Lozano:2020txg} AdS$_2$/CFT$_1$ duals with 4 supercharges were constructed using T-duality from the AdS$_3$/CFT$_2$ pairs  in  \cite{Lozano:2019emq,Lozano:2019zvg,Lozano:2019jza,Lozano:2019ywa}. 
These quivers  inherited some of the properties of the 2d ``mother'' CFTs, like the matter content that guarantees gauge anomaly cancelation in 2d. This condition does not have to be satisfied in one dimension, and indeed the quivers that we have constructed in this paper consist on a set of disconnected matrix models that do not satisfy it. It would be interesting to relate the SCQMs constructed in this paper to the quiver quantum mechanics  studied in 
\cite{Assel:2018rcw,Assel:2019iae}, living in D1-F1-D3-D5-NS5-D7 brane systems. These brane systems are the T-dual realisations of our D0-D4-D4'-D8-F1 brane configurations, so one would expect the dual SCQMs to be related.

 It was shown in  \cite{Faedo:2020nol} (see also \cite{Dibitetto:2018gtk}) that the AdS$_2\times S^3\times T_4\times I_{\rho}$ solutions with $h_4$ a particular function of the $T^4$ and $\rho$ asymptote locally to
 AdS$_6\times S^3\times I$ in the UV. This allowed to interpret these solutions as D0-D4-F1 line defect CFTs within the 5d $\mathcal{N}=1$ gauge theory living in the D4'-D8 brane system. It is plausible that we have found concrete realisations of these CFTs in the form of D0-D4 brane instantons interacting with F1 Wilson lines, connecting with the results in \cite{Tong:2014cha,Hwang:2014uwa,Chang:2016iji,Kim:2016qqs,Assel:2018rcw,Assel:2019iae}.
  
 We have seen that the holographic central charge can be related to the product of the electric and magnetic charges of the D-branes present in the solutions. This realises in a controlled string theory set-up the proposal in \cite{Hartman:2008dq}, where the central charge in the algebra of symmetry generators of AdS$_2$ with a gauge field was related to the square of the electric field, and adds to the controlled string theory set-ups realising this proposal already found in \cite{Dibitetto:2019nyz,Lozano:2020txg}. It would be interesting to see if further set-ups realising this proposal can be found in more general situations, especially in higher dimensions.
 
 Moreover, we have provided one more example where the holographic central charge can be derived from an action functional, following the ideas of geometric extremisation \cite{Couzens:2018wnk,Hosseini:2019ddy}.
Our results extend the results in \cite{Couzens:2018wnk,Hosseini:2019ddy} by the presence of sources and boundaries. We should stress that a physical reason that underlies the need for extremisation in a field theory with a non-Abelian R-symmetry, such as ours, remains as an interesting open problem that deserves further investigation.

An interesting open avenue that also deserves further investigation is the application of exact calculational techniques to our new backgrounds. This would allow to understand the properties of our theories in the IR and would set the stage for their applications to the study of black holes, along the lines of \cite{Benini:2015eyy}. It would be interesting to study our constructions with a more algebraic point of view, along the lines of
\cite{Fedoruk:2015lza}. Similarly, the connection or similarities with the dynamics uncovered in papers like \cite{Anninos:2013nra} should be nice to clarify.

\section*{Acknowledgements} We would like to thank Dionysios Anninos, Giuseppe Dibitetto,  Prem Kumar, Andrea Legramandi, Nicol\`o Petri, Guillermo Silva, Salomon Zacarias, for very useful discussions.
\\
Y.L. and A.R. are partially supported by the Spanish government grant
PGC2018-096894-B-100 and by the Principado de Asturias through the grant FC-GRUPINIDI/
2018/000174. AR is supported by CONACyT-Mexico.

\appendix 

\section{The seed backgrounds and their dual SCFTs}\label{AdS3}

In this Appendix we recall the main properties of the solutions to massive IIA supergravity (with localised sources) obtained in the recent work \cite{Lozano:2019emq}. It was proposed in \cite{Lozano:2019zvg}-\cite{Lozano:2019ywa} that these backgrounds are
holographic duals to two dimensional CFTs preserving ${\cal N}=(0,4)$ SUSY that we also summarise below. 
\\
The Neveu-Schwarz (NS) sector of these bosonic solutions reads,
\begin{align}
\text{d}s^2&= \frac{u}{\sqrt{\hat{h}_4 h_8}}\bigg(\text{d}s^2_{\text{AdS}_3}+\frac{h_8\hat{h}_4 }{4 h_8\hat{h}_4+(u')^2}\text{d}s^2_{S^2}\bigg)+ \sqrt{\frac{\hat{h}_4}{h_8}}\text{d}s^2_{\text{CY}_2}+ \frac{\sqrt{\hat{h}_4 h_8}}{u} \text{d}\rho^2,\label{kkktmba}\\
e^{-\Phi}&= \frac{h_8^{\frac{3}{4}} }{2\hat{h}_4^{\frac{1}{4}}\sqrt{u}}\sqrt{4h_8 \hat{h}_4+(u')^2},~~~~ H_{(3)}= \frac{1}{2}\text{d}(-\rho+\frac{ u u'}{4 \hat{h}_4 h_8+ (u')^2})\wedge\hat{\text{vol}}_{S^2}+ \frac{1}{h_8^2}\text{d}\rho\wedge H_2.\nn
\end{align}
Generically, the warping function $\hat{h}_4$ has support on $(\rho,\text{CY}_2)$. On the other hand,  $u$ and $h_8$ only depend of $\rho$. We denote $u'= \partial_{\rho}u$ and similarly for $h_8'$. 
The RR fluxes are 
\begin{subequations}
\begin{align}
F_{(0)}&=h_8',\;\;\;F_{(2)}=-\frac{1}{h_8}H_2-\frac{1}{2}\bigg(h_8- \frac{ h'_8 u'u}{4 h_8 \hat{h}_4+ (u')^2} \bigg)\hat{\text{vol}}_{S^2},\label{eq:classIflux2}\\[2mm]
F_{(4)}&=-\bigg(\text{d}\left(\frac{u u'}{2\hat{ h}_4}\right)+2 h_8  \text{d}\rho\bigg) \wedge\hat{\text{vol}}_{\text{AdS}_3}\nn\\[2mm]
& -\frac{h_8}{u} (\hat \star_4 \text{d}_4 \hat{h}_4)\wedge \text{d}\rho- \partial_{\rho}\hat{h}_4\hat{\text{vol}}_{\text{CY}_2}-\frac{u u'}{2 h_8( 4h_8 \hat{h}_4+ (u')^2)} H_2\wedge \hat{\text{vol}}_{S^2},\label{eq:classIflux3}
\end{align}
\end{subequations}
with the higher fluxes related to them as $F_{(6)}=-\star_{10} F_{(4)},~F_{(8)}=\star_{10} F_{(2)},~F_{(10)}=-\star_{10} F_{(0)}$.
It was shown in \cite{Lozano:2019emq} that supersymmetry holds whenever
\beq \label{susyC}
u''=0,~~~~ H_2+ \hat{\star}_4 H_2=0,
\eeq
where $\hat{\star}_4$ is the Hodge dual on CY$_2$. In what follows we will restrict ourselves to the set of solutions for which $H_2=0$
and
 $\hat{h}_4=\hat{h}_4(\rho)$.
 After this, the background reads,
\begin{eqnarray}
\text{d}s_{st}^2&=& \frac{u}{\sqrt{\hat{h}_4 h_8}}\bigg(\text{d}s^2_{\text{AdS}_3}+\frac{h_8\hat{h}_4 }{4 h_8 \hat{h}_4+(u')^2}\text{d}s^2_{S^2}\bigg)+ \sqrt{\frac{\hat{h}_4}{h_8}}\text{d}s^2_{\text{CY}_2}+ \frac{\sqrt{\hat{h}_4 h_8}}{u} \text{d}\rho^2,\nn \\
e^{-\Phi}&=& \frac{h_8^{\frac{3}{4}} }{2\hat{h}_4^{\frac{1}{4}}\sqrt{u}}\sqrt{4h_8 \hat{h}_4+(u')^2},~~~~~~ B_2= \frac{1}{2}\left(-\rho+ 2\pi k+\frac{ u u'}{4 \hat{h}_4 h_8+ (u')^2} \right) \hat{\text{vol}}_{S^2},\nn\\
\hat{F}_{(0)}&=&h_8',\;\;\;\;
\hat{F}_{(2)}=-\frac{1}{2}\bigg(h_8- h_8'(\rho-2\pi k)\bigg)\hat{\text{vol}}_{S^2},\nn\\
\hat{F}_{(4)}&=& -\bigg(\partial_\rho\left(\frac{u u'}{2 \hat{h}_4}\right)+2 h_8\bigg)  \text{d}\rho \wedge\hat{\text{vol}}_{\text{AdS}_3}
- \partial_{\rho}\hat{h}_4\hat{\text{vol}}_{\text{CY}_2}.\label{eq:background}
\end{eqnarray}
We have written the Page fluxes, $\hat{F}=e^{-B_{(2)}}\wedge F$, more useful for our purposes. Notice that we have also allowed for large gauge transformations $B_{(2)}\to B_{(2)} + { \pi k}\; \hat{\text{vol}}_{S^2}$, for $k=0,1,...., P$. The transformations are performed every time we cross an interval $[2\pi k, 2\pi(k+1)]$. The space begins at $\rho=0$ and ends at $\rho=2\pi(P+1)$. This will become more apparent once the functions $\hat{h}_4, h_8, u$ are specified below.


The background in \eqref{eq:background} is a SUSY solution  of the massive IIA equations of motion if the functions $\hat{h}_4,h_8,u$ satisfy (away from localised sources),
\begin{equation}
\hat{h}_4''(\rho)=0,\;\;\;\; h_8''(\rho)=0,\;\;\;\; u''(\rho)=0.\label{eqsmotion}
\end{equation}
Various particular solutions were analysed in \cite{Lozano:2019emq}. Here we will  consider an infinite family of solutions for which the functions are piecewise continuous.  These were carefully studied in \cite{Lozano:2019zvg}-\cite{Lozano:2019ywa} and a precise dual field theory was proposed. Let us briefly summarise aspects of these SCFTs.

\subsection*{The associated dual SCFTs}
A generic background of the form in \eqref{eq:background} is defined by the functions $\hat{h}_4,h_8, u$.  For the type of solutions  that were considered in 
 \cite{Lozano:2019zvg}-\cite{Lozano:2019ywa} the $\rho$-interval was bounded in $[0, 2\pi(P+1)]$. This range was determined by the vanishing of the functions $\hat{h}_4, h_8$.
Generically these functions are piecewise linear in $\rho$, and can be taken as,
 \begin{equation} \label{profileh4sp2}
\hat{h}_4(\rho)\!=\!\Upsilon\! \,h_4(\rho)\!=\!\!
                    \Upsilon\!\!\,\left\{ \begin{array}{ccrcl}
                       \frac{\beta_0 }{2\pi}
                       \rho & 0\leq \rho\leq 2\pi \\
                 \alpha_1+\frac{\beta_1}{2\pi}(\rho-2\pi)      &2\pi \leq \rho \leq 4\pi \\
\alpha_2+\frac{\beta_2}{2\pi} (\rho-4\pi) & 4\pi\leq\rho\leq 6\pi\\
                                     \alpha_k\! +\! \frac{\beta_k}{2\pi}(\rho-2\pi k) &~~ 2\pi k\leq \rho \leq 2\pi(k+1),\;\;\;\; k:=3,....,P-1\\
                      \alpha_P-  \frac{\alpha_P}{2\pi}(\rho-2\pi P) & 2\pi P\leq \rho \leq 2\pi(P+1).
                                             \end{array}
\right.
\end{equation}
 \begin{equation} \label{profileh8sp2}
h_8(\rho)
                    =\left\{ \begin{array}{ccrcl}
                       \frac{\nu_0 }{2\pi}
                       \rho & 0\leq \rho\leq 2\pi \\
                 \mu_1+\frac{\nu_1}{2\pi}(\rho-2\pi)      &2\pi \leq \rho \leq 4\pi \\
\mu_2+\frac{\nu_2}{2\pi} (\rho-4\pi) & 4\pi\leq\rho\leq 6\pi\\
                                     \mu_k+ \frac{\nu_k}{2\pi}(\rho-2\pi k) &~~ 2\pi k\leq \rho \leq 2\pi(k+1),\;\;\;\; k:=3,....,P-1\\
                      \mu_P-  \frac{\mu_P}{2\pi}(\rho-2\pi P) & 2\pi P\leq \rho \leq 2\pi(P+1),
                                             \end{array}
\right.
\end{equation}
where 
\begin{equation}
\alpha_k=\sum_{j=0}^{k-1}\beta_j\, , \qquad \mu_k=\sum_{j=0}^{k-1}\nu_j\, .
\end{equation}
The choice of constants is imposed by continuity of the metric and dilaton, while the fluxes can present discontinuities associated to the presence of branes (see \cite{Lozano:2019emq} for more details).
In turn, the $u$ function, also linear in $\rho$, does not enter in the magnetic Page fluxes associated to the solutions, and thus, does not affect the type of quivers that can be constructed from the brane set-up. Here we will restrict to the simplest case, $u'=0$.

\begin{figure}[h!]
    \centering
    {{\includegraphics[width=10cm]{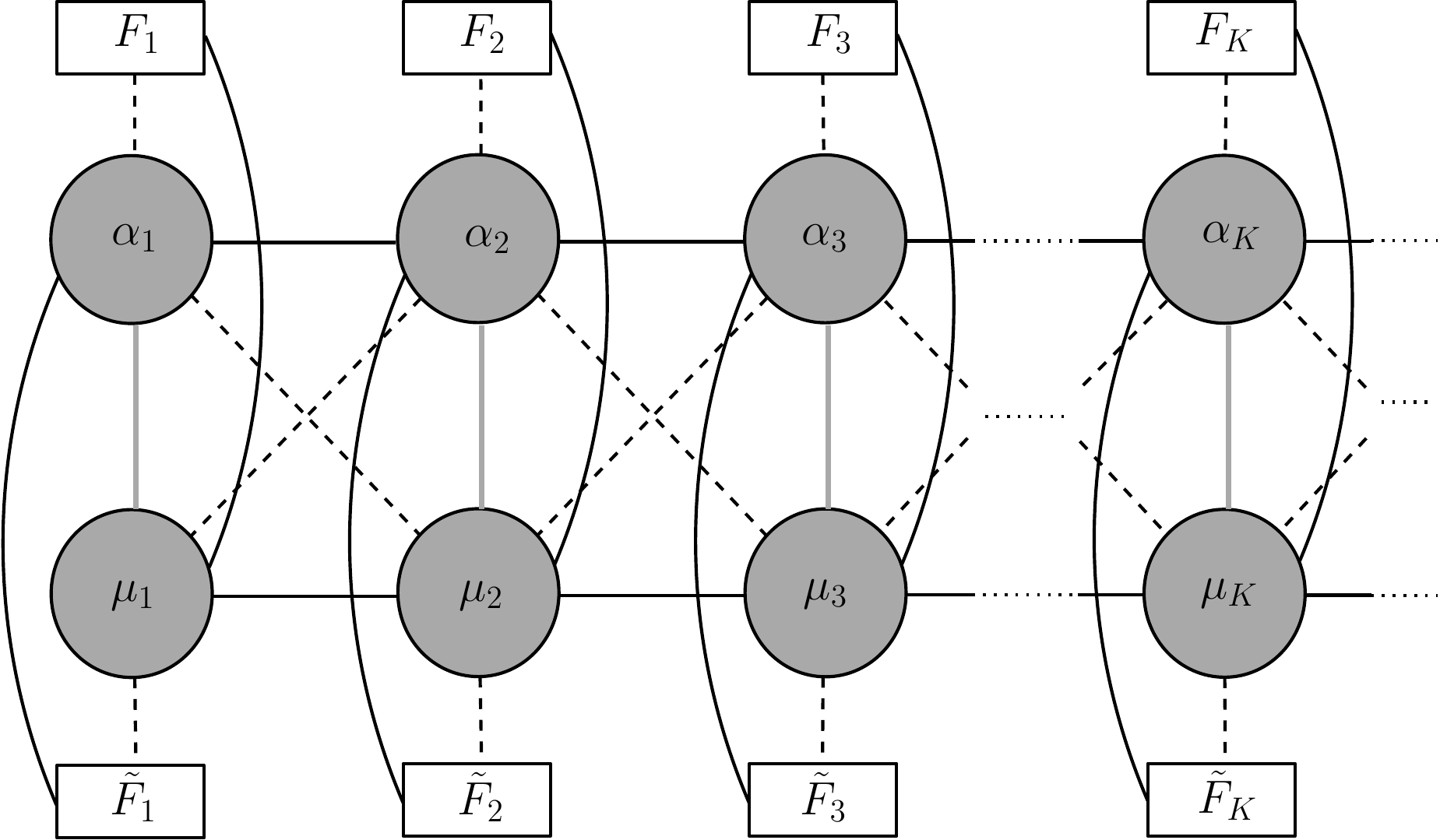} }}%
   \caption{ A generic quiver field theory whose IR is dual to the holographic background defined by the functions in \eqref{profileh4sp2}-\eqref{profileh8sp2}. The solid black line represents a $(4,4)$ hypermultiplet. The grey line represents a $(0,4)$ hypermultiplet and  the dashed line represents a $(0,2)$ Fermi multiplet. ${\cal N}=(4,4)$ vector multiplets are the degrees of freedom  in each gauged node.}
\label{figurageneral}
\end{figure}

The background in eq.\eqref{eq:background} for the functions  $\hat{h}_4,h_8$ specified above is dual to the CFT describing the low energy dynamics of a two dimensional quantum field theory encoded by the  quiver in Figure \ref{figurageneral}.
In this quiver the ranks of the flavour groups are determined by the absence of gauge anomalies, to be  \cite{Lozano:2019zvg}-\cite{Lozano:2019ywa}
\begin{equation}
F_k=\nu_{k-1}-\nu_{k},\;\;\;\; \tilde{F}_k=\beta_{k-1}-\beta_{k}.
\end{equation}
The most stringent check in \cite{Lozano:2019zvg}-\cite{Lozano:2019ywa}  for the validity of this proposal was the matching between the field theory and holographic central charges. The $U(1)_R$ 
current two-point function can be identified when flowing to the far IR (a conformal point is reached) with the right moving central charge of the ${\cal N}=(0,4)$ conformal field theory. Following \cite{Tong:2014yna}, \cite{Putrov:2015jpa} it was found  for a generic quiver with $n_{hyp}$ hypermultiplets and $n_{vec}$ vector multiplets,  that the central charge (the anomaly of the $U(1)_R$ current) is
\begin{equation}
c_{CFT}=6 (n_{hyp}- n_{vec}).\label{centralcft}
\end{equation}
In  \cite{Lozano:2019zvg}-\cite{Lozano:2019ywa} a variety of examples of long linear quivers for which the ranks of each of the nodes is a large number were discussed. In each of these qualitatively different examples, it was found that the field theoretical central charge of eq.(\ref{centralcft})
coincides (in the limit of long quivers with large ranks) with the holographic central charge, computed as
\begin{equation}
c_{hol}= \frac{3\pi}{2 G_N}\text{Vol}_{\text{CY}_2}\int_0^{2\pi (P+1)} \hat{h}_4 h_8 \, \text{d} \rho= \frac{3}{\pi} \int_0^{2\pi (P+1)} {h}_4 h_8 \, \text{d} \rho  ,\label{chol}
\end{equation}
where $G_N= 8\pi^6$ (with $g_s=\alpha'=1$) and $\Upsilon \text{Vol}_{\text{CY}_2}=16\pi^4$.

\section{D0-D4-D4'-D8 quantum mechanics} \label{open-strings}

In this appendix we include a study of the low-energy field content emerging from the brane web given in Table \ref{Table brane web set type IIA3}. The four coordinates $x^{1,2,3,4}$ parametrise a Calabi-Yau, whose nature we need not specify in general. Just to be concrete we can take it to be a four-torus. 
The compactification to a four-torus breaks the SO$(4)$ symmetry associated to rotations on the four-dimensional subspace spanned by $x^{1,2,3,4}$. However, in what follows, we will continue to use the SO$(4)$ algebra to organise fields\footnote{See \cite{David:2002wn}, Ch. 4, where subtleties and special limits concerning quantisation on a compact four-torus are discussed.}.

As all the branes are localised in the $x^5 = \rho$ direction, strings stretched between branes in adjacent intervals $\mathcal{I}_k$ and $\mathcal{I}_{k+1}$, with $\mathcal{I}_k = [2 \pi k, 2 \pi (k+1)]$, are necessarily long, hence describing massive states. Therefore, the Hilbert space of the full system is given by the direct sum of individual Hilbert spaces associated with D0-D4-D4'-D8 systems, that we now describe.

In the following, we will use that the rank of the gauge and flavour groups associated with D0, D4, D4' and D8 branes in a given interval $\mathcal{I}_k$ is given by $\mu_k$, $\alpha_k$, $\tilde{F}_k$ and $F_k$, respectively, with $\tilde{F}_k = \beta_{k-1} - \beta_k$ and $F_k = \nu_{k-1} - \nu_k$, in agreement with the notation used in the main text. The elementary excitations on the branes are given by strings with ends attached to the branes.
We will often use 2d language, exploiting the fact that our brane web is T-dual to a D1-D5-D5'-D9 system, and we need only to dimensionally reduce to 0 + 1 dimensions to get back to D0-D4-D4'-D8.
Much of the conventions in denoting fields is borrowed from \cite{Tong:2014yna}, while quantisation of open strings can be easily done following \cite{Polchinski:1998rq}, to which we refer for more details.

\begin{itemize}

\item D0$-$D0 strings: For simplicity, let us start by considering a sytem of $\mu_k$ parallel D1$-$D1 branes stretched along $x^{0, 5}$ (eventually we will reduce down to 1d). The system consists of a $\text{U}(\mu_k)$ gauge theory and can be obtained by dimensional reduction of a 10d $\mathcal{N} = 1$ $\text{U}(\mu_k)$ gauge theory, where the field content is that of a 10d gauge field and a 10d Majorana-Weyl spinor.

In order to see what this entails, consider the decomposition of the 10d Lorentz group as $\text{SO}(1,9) \rightarrow \text{SO}(1,1) \times \text{SO}(8)$. A Majorana-Weyl spinor in the $\mathbf{16}$ of\footnote{We should really say Spin$(1,9)$. In this appendix, we will not care much about global structure of groups.} SO$(1,9)$ can be decomposed as $\mathbf{16} = (+\frac{1}{2}, \mathbf{8}_s) \oplus (- \frac{1}{2}, \mathbf{8}_c)$. We should then further split the $\text{SO}(8)$ as $\text{SO}(8) \rightarrow G$ with $G = \text{SO}(4)^- \times \text{SO}(4)^+$, where the two SO$(4)$'s are realised on the D4 and D4' branes. We may also rewrite $G$ as
\begin{equation}
G = \text{SU}(2)_L^- \times \text{SU}(2)_R^- \times \text{SU}(2)_L^+ \times \text{SU}(2)_R^+ \, ,
\end{equation}
where $\text{SU}(2)_R^- \times \text{SU}(2)_R^+$ can be understood as the R-symmetry of the quantum mechanical system.

A ten-dimensional gauge field decomposes as a two-dimensional gauge field $u_{\mu}$ plus $8$ scalars. The scalars, that we denote collectively as $Y$ and $Z$, transform in the $(\mathbf{1}, \mathbf{1}, \mathbf{2}, \mathbf{2})$ and $(\mathbf{2}, \mathbf{2}, \mathbf{1}, \mathbf{1})$ of $G$, respectively. Decomposing also the eight-dimensional spinors under representations of $G$, and using $\mathcal{N} = (0,4)$ language, we get\footnote{We are considering the following decomposition
\begin{equation}
\mathbf{8}_s = (\mathbf{2}, \mathbf{1}, \mathbf{1}, \mathbf{2}) \oplus (\mathbf{1}, \mathbf{2}, \mathbf{2}, \mathbf{1}) \, , \quad 
\mathbf{8}_c = (\mathbf{2}, \mathbf{1}, \mathbf{2}, \mathbf{1}) \oplus (\mathbf{1}, \mathbf{2}, \mathbf{1}, \mathbf{2}) \, ,
\end{equation}
from which \eqref{(8, 8) vectormultiplet} follows. The eight complex fermions obtained by decomposing a 10d Majorana-Weyl spinor are denoted, in some compact notation, as $\zeta$, $\tilde{\zeta}$, $\lambda$ and $\tilde{\lambda}$.
}
\begin{equation}\label{(8, 8) vectormultiplet}
\begin{split}
\text{Vector multiplet:} \quad  & u_{\mu} \quad (\mathbf{1}, \mathbf{1}, \mathbf{1}, \mathbf{1}) \\
&\zeta \quad (\mathbf{1}, \mathbf{2}, \mathbf{1}, \mathbf{2}) \\
\text{Twisted hypermultiplet:} \quad  &Y \quad (\mathbf{1}, \mathbf{1}, \mathbf{2}, \mathbf{2}) \\
&\tilde{\zeta} \quad (\mathbf{1}, \mathbf{2}, \mathbf{2}, \mathbf{1}) \\
\text{Hypermultiplet:} \quad  &Z \quad (\mathbf{2}, \mathbf{2}, \mathbf{1}, \mathbf{1}) \\
&\lambda \quad (\mathbf{2}, \mathbf{1}, \mathbf{1}, \mathbf{2}) \\
\text{Fermi multiplet:} \quad  & \tilde{\lambda}  \quad (\mathbf{2}, \mathbf{1}, \mathbf{1}, \mathbf{2})
\end{split}
\end{equation}
i.e. the field content of a $(4, 4)$ vector and a $(4 ,4)$ hypermultiplet. The latter was denoted in the main text as a black solid line starting and ending on the same gauge groups, see Figure \ref{quivers}. All the multiplets above transform in the adjoint of the gauge group U($\mu_k$) on the D1 branes.

Let us now dimensionally reduce to one-dimensional field theory.
The gauge field $u_{\mu}$, upon dimensional reduction, decomposes as $u_{\mu} \rightarrow (u_t, \sigma)$, with $u_t$ the 1d vector field and $\sigma$ a real scalar of the 1d theory. The fermions and scalars are inert, leading indeed to the field content \eqref{(8, 8) vectormultiplet}. Of course, going to down to one dimension, chirality for the fermions is lost.

Quantisation of D4$-$D4 strings, upon reduction to 0+1 dimensions, leads to the same conclusions for the gauge group U$(\alpha_k)$.

\item $\text{D0} - \text{D4}'$ strings: The system of D0$-$D4' branes is a brane web with 4 ND relative boundary conditions, T-dual of the well-known D1$-$D5 system.
The NS and R sectors give rise, upon imposing the GSO projection, to a scalar in the $\mathbf{(1, 2)}$ of the internal SO$(4)^+$ and a 6d Weyl spinor which is a singlet under the internal SO$(4)^+$. We can dimensionally reduce to 1d
to get an $\mathcal{N}=(4,4)$ twisted hypermultiplet. In $\mathcal{N}=(0, 4)$ language we have
\begin{equation}\label{hyper D1-D5}
\begin{split}
\text{Twisted Hypermultiplet:} \quad  &\phi' \quad (\mathbf{1}, \mathbf{1}, \mathbf{1}, \mathbf{2}) \\
&\psi' \quad (\mathbf{1}, \mathbf{2}, \mathbf{1}, \mathbf{1}) \\
\text{Fermi multiplet:} \quad  &\tilde{\psi}'\quad (\mathbf{2}, \mathbf{1}, \mathbf{1}, \mathbf{1})
\end{split}
\end{equation}

Each of these multiplets transform in the $\boldsymbol{\mu_k}$ of the gauge group U$(\mu_k)$ and in the anti-fundamental of the global SU$(\tilde{F}_k)$.

\item $\text{D0} - \text{D4}$ strings: The field content is the same as for the D0$-$D4' system. The only difference is that $\text{SO}(4)^- \times \text{SO}(4)^+$ are exchanged. From our discussion about D0 $-$ D4' strings we find
\begin{equation}
\begin{split}
\text{Hypermultiplet:} \quad  &\tilde{\phi} \quad (\mathbf{1}, \mathbf{2}, \mathbf{1}, \mathbf{1}) \\
&\psi \quad (\mathbf{1}, \mathbf{1}, \mathbf{1}, \mathbf{2}) \\
\text{Fermi multiplet:} \quad  &\tilde{\psi} \quad (\mathbf{1}, \mathbf{1}, \mathbf{2}, \mathbf{1})
\end{split}
\end{equation}
i.e. a $(4, 4)$ hypermultiplet. Each of these multiplets transform in the $\boldsymbol{\mu_k}$ of the gauge group U$(\mu_k)$ and in the $\boldsymbol{\overline{\alpha}_k}$ of the other gauge group U$(\alpha_k)$.

\item $\text{D4} - \text{D4'}$ strings: This system is an example of a brane web with $8$ relative ND boundary conditions. Strings in the NS sector are automatically massive. Indeed, such a brane web can be T-dualised to a $\text{D0} - \text{D8}$ system, where the R sector gives rise to an $\mathcal{N} = (0,2)$ Fermi multiplet, singlet of all internal symmetries,
\begin{equation}
\mathcal{N} = (0,2) \quad \text{Fermi multiplet:} \quad \chi \quad (\mathbf{1}, \mathbf{1}, \mathbf{1}, \mathbf{1}) \, .
\end{equation}

Such a Fermi multiplet transforms in the $\boldsymbol{\alpha_k}$ of the gauge group $\text{U}(\alpha_k)$ and in the antifundamental of the global group $ \text{SU}(\tilde{F}_k)$.

\item $\text{D0} -\text{D8}$ strings: Again, this is a system with $8$ relative ND boundary conditions. Therefore, also in this case we find a $\mathcal{N} = (0,2)$ Fermi multiplet, which is a singlet of all internal symmetries
\begin{equation}
\mathcal{N} = (0,2) \quad  \text{Fermi multiplet:} \quad \xi \quad (\mathbf{1}, \mathbf{1}, \mathbf{1}, \mathbf{1}) \, .
\end{equation}

This $\mathcal{N} = (0,2)$ Fermi multiplet transforms in the $\boldsymbol{\mu_k}$ of the gauge group $\text{U}(\mu_k)$ and in the $\boldsymbol{\overline{F}_k}$ of the global group $ \text{SU}(F_k)$.

\item $\text{D4} -\text{D8}$ strings: Finally, we have a system of D4 $-$ D8 branes with $4$ relative ND boundary conditions, with degenerate NS and R sectors. The quantisation of D4 $-$ D8 strings leads again to an $\mathcal{N}=(4,4)$ hypermultiplet. Its field content is given by
\begin{equation}\label{(4, 4) hyper}
\begin{split}
\text{Hypermultiplet:} \quad  &\varphi \quad (\mathbf{1}, \mathbf{1}, \mathbf{1}, \mathbf{2}) \\
&\eta \quad (\mathbf{1}, \mathbf{2}, \mathbf{1}, \mathbf{1}) \\
\text{Fermi multiplet:} \quad  &\tilde{\eta} \quad (\mathbf{2}, \mathbf{1}, \mathbf{1}, \mathbf{1}) \, ,
\end{split}
\end{equation}
where each multiplet transforms in the $\boldsymbol{\alpha_k}$ of the gauge group $\text{U}(\alpha_k)$ and in the $\boldsymbol{\overline{F}_k}$ of the global group $ \text{SU}(F_k)$.

\end{itemize}

Putting altogether, we get the field content of Figure \ref{quivers}, where black solid lines represent $(4, 4)$ hypermultiplets and dashed black lines $(0, 2)$ Fermi multiplets. Interactions between the various fields can be constructed following the rules of e.g. \cite{Tong:2014yna, Tong:2014cha}. In particular, interactions should not spoil supersymmetry. See \cite{Tong:2014yna} where this is done consistently.

\section{Holographic calculation of SQM observables}

In this appendix we probe the background in \eqref{NS sector analytically continued} and \eqref{RR sector analytically continued} by introducing probe D branes. The action describing the coupling of a generic Dp brane to the NS-NS and RR closed string fields contains the usual DBI $+$ WZ terms,
\begin{equation}
\begin{split}
S_{\text{Dp}} &= S_{DBI} + S_{WZ} \, , \\
S_{DBI} &= - T_{\text{p}} \int \text{d}^{p+1} \xi \left\{ e^{-\phi} \left[ - \det (g_{ab} + B_{ab} + 2 \pi \alpha' \mathcal{F}_{ab}) \right]^{\frac{1}{2}}  \right\} \\
S_{WZ}&= T_{\text{p}} \int_{p+1} \exp (2 \pi \alpha' \mathcal{F}_{(2)} + B_{(2)}) \wedge \sum_{q} C_{(q)} \, , \\
\end{split}
\end{equation}
where $\mathcal{F}$ is the gauge field living on the brane.
\begin{itemize}

\item Let us begin by considering probe D0 branes. The field theory living on a D0 brane is $(0+1)-$dimensional. We can define a metric for such a $(0+1)-$dimensional field theory from the pullback of \eqref{NS sector analytically continued},
\begin{equation}
\text{d}s^2_{\text{ind}} = - \frac{u \sqrt{\hat{h}_4 h_8}}{4 \hat{h}_4 h_8 - (u')^2} \text{d} t^2 \, .
\end{equation}
Note that the pullback of $B_{(2)}$ and the field strength $\mathcal{F}_{ab}$ on the D0 brane are automatically zero, due to their (anti)symmetric properties. Given that
\begin{equation}
e^{-\Phi} \sqrt{\det g_{\text{ind}}} = \frac{h_8(\rho_*)}{2} \cosh (r_{*}) \, ,
\end{equation}
where the ``$*$" simply refers to the fact that we are keeping $r$ and $\rho$ fixed at some values, for a probe D0 brane we find the following action
\begin{equation}
S_{\text{D0}} = - T_{\text{0}} \int_{\mathbb{R}} \text{d} t \,e^{-\Phi} \sqrt{\det g_{\text{ind}}} + T_{\text{0}} \int C_{(1)} \, ,
\end{equation}
which leads to
\begin{equation}
S_{\text{D0}} = - T_{\text{0}} \frac{h_8(\rho_*)}{2} \cosh (r_{\star}) \int_{\mathbb{R}} \text{d} t + T_{\text{0}} \frac{\mu_k}{2} \sinh (r_{\star}) \int_{\mathbb{R}} \text{d} t \, .
\end{equation}
If we choose $\rho_* = 2 \pi k$, we find
\begin{equation}
S_{\text{D0}} = T_{\text{0}} \frac{\mu_k}{2} ( \sinh r_* - \cosh r_*) \int_{\mathbb{R}} \text{d} t \, .
\end{equation}
We then find that the brane is calibrated only when $r_* = \infty$, for which, however, we have a vanishing action.

Consider instead the coupling of a D0 brane to an external one form, call it $\mathcal{A}_{(1)}$. The action describing such a coupling is
\begin{equation}
S = 2 \pi T_{\text{0}} \int F_{(0)} \mathcal{A}_{(1)} \, .
\end{equation}
If, as in \eqref{RR sector analytically continued}, $F_{(0)} = h_8'$, we get
\begin{equation}\label{action D0 CS}
S = 2 \pi T_{\text{D0}} \frac{\nu_k}{2 \pi} \int_{\mathbb{R}} \mathcal{A}_{(t)} \text{d} t \, .
\end{equation}
Using that $T_{\text{0}} = 1$, we find that the action \eqref{action D0 CS} describes a Chern-Simons term with $k_{CS} = \nu_k$. If we were to make sense of the action \eqref{action D0 CS} also on a topologically non-trivial one-dimensional manifold, say on a circle $S^1$, we would find that it describes a gauge invariant action if and only if $\nu_k$ is an integer. Moreover, without loss of generality, we can take $\nu_k$ to be positive as well, as a Chern-Simons term is odd under parity.

\item Let us move on to the case of a probe D4 brane stretched along $(t, \text{CY}_2)$. The induced metric on the worldvolume of such a D4 brane, from the pullback of \eqref{NS sector analytically continued}, reads
\begin{equation}
\text{d}s^2_{\text{ind}} = - \frac{u \hat{h}_4 h_8}{4 \hat{h}_4 h_8 - (u')^2} \text{d} t^2 + \sqrt{\frac{\hat{h}_4}{h_8}} \text{d}s^2_{\text{CY}_2} \, .
\end{equation}
The pullback of $B_{(2)}$ vanishes, whereas we will ignore for now the field strength $\mathcal{F}_{ab}$ along the brane\footnote{An $\alpha'$ expansion of the DBI action would produce a Maxwell kinetic term for $\mathcal{F}$. Eventually we will be interested in the dimensional reduction to 1 dimension where such a kinetic term would be absent.}. Given that
\begin{equation}
e^{-\Phi} \sqrt{\det g_{\text{ind}}} = \frac{\hat{h}_4(\rho_*)}{2} \cosh (r_{*}) \, ,
\end{equation}
for a probe colour D4 brane the DBI action reads
\begin{equation}
S_{DBI} = - T_{4} \text{Vol}_{\text{CY}_2} \frac{\hat{h}_4(\rho_*)}{2} \cosh (r_{*}) \int_{\mathbb{R}} \text{d} t \, .
\end{equation}
Using that $T_{\text{4}} = 1/(2 \pi)^4$ and $\hat{h}_4 = \Upsilon h_4$ we find
\begin{equation}
S_{DBI} = - \frac{\Upsilon \text{Vol}_{\text{CY}_2}}{(2 \pi)^4} \frac{h_4(\rho_*)}{2} \cosh (r_{*}) \int_{\mathbb{R}} \text{d} t \, .
\end{equation}

Let us consider now the WZ part of the D4 brane action
\begin{equation}
S_{WZ} = T_{\text{4}} \int C_{(5)} + 2 \pi C_{(3)} \wedge \mathcal{F}_{(2)} + 4 \pi^2 C_{(1)} \wedge \mathcal{F}_{(2)} \wedge \mathcal{F}_{(2)} \, .
\end{equation}

If we place the D4 brane at $\rho = 2 \pi k$, its action reads 
\begin{equation}
\begin{split}
S &= S_{DBI} + T_{\text{4}} \int C_{(5)} + 2 \pi T_{\text{4}} \int \hat{F}_{(4)} \wedge \mathcal{A}_{(1)} \, , \\
&= T_{\text{4}} \Upsilon \text{Vol}_{\text{CY}_2} \frac{\alpha_k}{2} ( \sinh r_* - \cosh r_*) \int_{\mathbb{R}} \text{d} t + 2 \pi T_{\text{4}} \int \hat{F}_{(4)} \wedge \mathcal{A}_{(1)} \, .
\end{split}
\end{equation}
Here we have used that  $C_{(5)}\rvert_{\text{D4}} = - \frac{1}{2} (\hat{h}_4 - \hat{h}_4' (\rho - 2 \pi k)) \sinh(r) \text{d}t \wedge \hat{\text{vol}}_{\text{CY}_2}$, where $\text{d} C_{(5)} = \hat{F}_{(6)}$ on the brane. In turn, the term $2 \pi C_{(3)} \wedge F_{(2)}$ contributes as a Chern-Simons term, $2 \pi \hat{F}_{(4)} \wedge \mathcal{A}_{(1)}$, with 
$\hat{F}_{(4)} = \hat{h}_4' \hat{\text{vol}}_{\text{CY}_2}$. This Chern-Simons term reads explicitly,
\begin{equation}
2 \pi T_{\text{4}} \int \hat{F}_{(4)} \wedge \mathcal{A}_{(1)} = 2 \pi T_{\text{4}} \hat{h}_4' \text{Vol}_{\text{CY}_2} \int \mathcal{A}_{(1)} \, .
\end{equation}
Using that $\hat{h}_4 = \Upsilon h_4$ and choosing $\Upsilon$ as usual such that $\Upsilon T_{\text{4}} \text{Vol}_{\text{CY}_2} = 1$ we find a CS action with level $k_{CS} = \beta_k$. Again, invariance under large gauge transformations implies that $\beta_k$ has to be an integer. Parity allows us to take it to be positive.

\end{itemize}

\section{Continuity of the NS sector}

\paragraph{} Let us comment briefly on the continuity of the NS sector of \eqref{NS sector analytically continued}. Being $u$, $\hat{h}_4$ and $h_8$ linear functions of $\rho$, we consider the following expressions for them in the interval $[2 \pi k, 2 \pi (k+1)]$,
\begin{equation}
\hat{h}_4^{(k)} = \alpha_k + \frac{\beta_k}{2 \pi} (\rho - 2 \pi k) \, , \quad h_8^{(k)} = {\mu}_k + \frac{{\nu}_k}{2 \pi} (\rho - 2 \pi k) \, , \quad u^{(k)} = a_k + \frac{b_k}{2 \pi} (\rho - 2 \pi k) \, .
\end{equation}

Let us now see what conditions should be imposed on the constants $\{ \alpha_k, \beta_k, \mu_k, \nu_k, a_k, b_k \}$ in order for the NS sector in \eqref{NS sector analytically continued} to be continuous. We rewrite \eqref{NS sector analytically continued} as
\begin{equation}
\mathrm{d}s^2 = f_1 \text{d}s^2_{\text{AdS}_2} + f_2 \text{d}s^2_{S^3} + f_3 \text{d}s^2_{\text{CY}_2} + f_4 \text{d} \rho^2 \, , \quad \quad B_{(2)} = f_5 \hat{\text{vol}}_{\text{AdS$_2$}} \, , \quad \quad e^{-2 \Phi} = f_6 \, .
\end{equation}
Then, we should impose the continuity of the $f_i$'s at all points $\rho_{k} = 2 \pi k$, where the functions $u, \hat{h}_4$ and $h_8$ change defining laws. This is achieved by demanding
\begin{equation}\label{limit for continuity NS}
\lim_{\rho \rightarrow \rho_{k}^-} f_i = \lim_{\rho \rightarrow \rho_{k}^+} f_i \, .
\end{equation} 
We will refer to the left and right hand side of \eqref{limit for continuity NS} as $f_i^-$ and $f_i^+$, respectively, and therefore continuity corresponds to requiring $f_i^- = f_i^+$. A straightforward computation shows that
\begin{equation}
f_1^- = f_1^+ \Rightarrow \frac{(a_{k-1} + b_{k-1}) \sqrt{(\alpha_{k-1} + \beta_{k-1}) (\mu_{k-1} + \nu_{k-1})}}{16 \pi^2 (\alpha_{k-1} + \beta_{k-1}) (\mu_{k-1} + \nu_{k-1}) - b_{k-1}^2} = \frac{a_k \sqrt{\alpha_k \mu_k}}{16 \pi^2 \alpha_k \mu_k - b_k^2} \, ,
\end{equation}
\begin{equation}
f_2^- = f_2^+ \Rightarrow \frac{(a_{k-1} + b_{k-1}) }{\sqrt{(\alpha_{k-1} + \beta_{k-1}) (\mu_{k-1} + \nu_{k-1})}} = \frac{a_k}{\sqrt{\alpha_k \mu_k}} \, ,
\end{equation}
\begin{equation}
f_3^- = f_3^+ \Rightarrow \sqrt{\frac{\alpha_{k-1} + \beta_{k-1}}{\mu_{k-1} + \nu_{k-1}}} = \sqrt{\frac{\alpha_k}{\mu_k}}\, ,
\end{equation}
\begin{equation}
f_4^- = f_4^+ \Rightarrow \frac{\sqrt{(\alpha_{k-1} + \beta_{k-1}) (\mu_{k-1} + \nu_{k-1})}}{a_{k-1} + b_{k-1}} = \frac{\sqrt{\alpha_k \mu_k}}{a_{k}} \, ,
\end{equation}
\begin{equation}
f_5^- = f_5^+ \Rightarrow \frac{a_{k-1} b_{k-1} + 16 \pi^2 (\alpha_{k-1} + \beta_{k-1}) (\mu_{k-1} + \nu_{k-1})}{16 \pi^2 (\alpha_{k-1} + \beta_{k-1}) (\mu_{k-1} + \nu_{k-1}) - b_{k-1}^2} = -1 + \frac{a_k b_k}{b_k^2 - 16 \pi^2 \alpha_k \mu_k}\, ,
\end{equation}
\begin{equation}
f_6^- = f_6^+ \Rightarrow \frac{(\mu_{k-1} + \nu_{k-1})^{3/2} \left[ 16 \pi^2 (\alpha_{k-1} + \beta_{k-1}) (\mu_{k-1} + \nu_{k-1}) - b_{k-1}^2 \right]}{\sqrt{\alpha_{k-1} \beta_{k-1}} (a_{k-1} + b_{k-1})} = \frac{\mu_k^{3/2} \left[ 16 \pi^2 (\alpha_{k} + \beta_{k}) -b_k^2 \right]}{ a_{k} \sqrt{\alpha_k} } \, .
\end{equation}

A possible solution for such a system is
\begin{equation}
a_{k} = a_{k-1} + b_{k-1} \, , \quad b_{k} = b_{k-1} = b_0\, , \quad \alpha_{k} = \alpha_{k-1} - \beta_{k-1}\, , \quad \mu_{k} = \mu_{k-1} - \nu_{k-1} \, ,
\end{equation}
which, in turn, implies
\begin{equation}
a_{k} = a_0 + k b_0 \, , \quad \alpha_{k} = \alpha_0 + \sum_{j=0}^{k-1} \beta_j \, , \quad \mu_{k} = \mu_0 + \sum_{j=0}^{k-1} \nu_j \, .
\end{equation}
These are the very same conditions assuring continuity of $u$, $\hat{h}_4$ and $h_8$. Therefore, we conclude that imposing continuity of the functions $u$, $\hat{h}_4$ and $h_8$ is sufficient to get continuity for the NS sector.

\section{Volumes and stringy volumes}\label{appendix1}

Analysing the volume of the compact submanifolds of the solutions in eqs. (\ref{profileh4sp})-(\ref{profileh8sp}) we run into the possibility that some of these submanifolds have infinite size. However, in spite of a divergent warp factor, the ``stringy size'' of the submanifold is actually finite or vanishing at the ends of the space. The finite stringy-volume case does not pose any problem in interpreting a D-brane wrapping such cycle. The case in which the cycle shrinks may suggest an interpretation of the singularity in terms of new massless degrees of freedom (branes wrapping the shrinking cycles) that the supergravity solution is not encoding.

\begin{figure}
\centering
\includegraphics[scale=0.73]{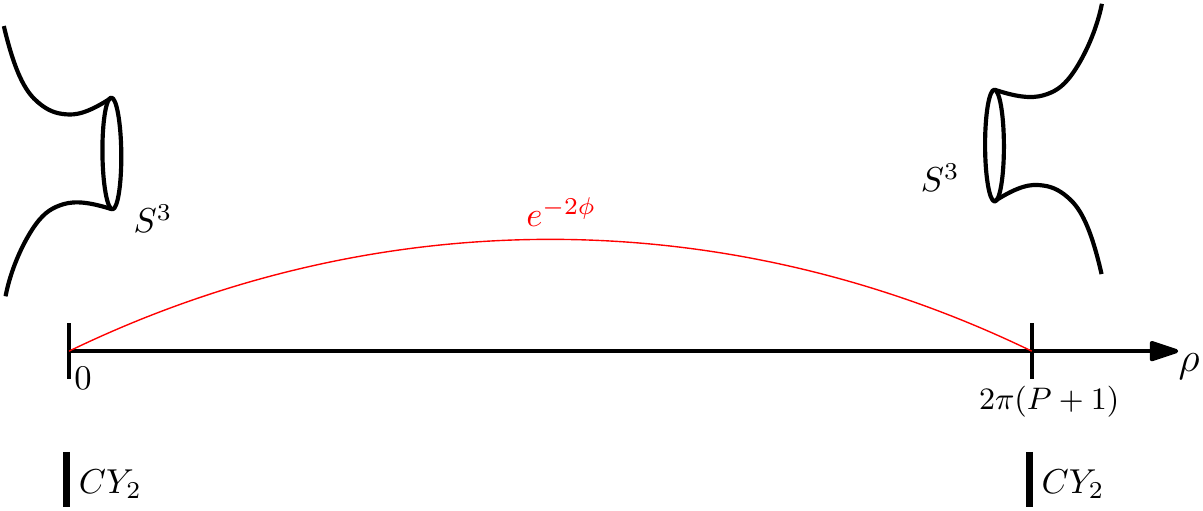}
\caption{Behaviour of the solutions at both ends of the $\rho$-interval for $u=$ constant. We find that the CY$_2$ is of finite size, while the $S^3$ diverges at both ends of the interval.}
\label{figuraC1}
\end{figure}

In the case with $u'=0$ we see in eqs.(\ref{forAppC1}), (\ref{forAppC2}) that the $S^3$ diverges at both ends of the $\rho$-interval. A representation of the various compact submanifolds is 
given in Figure \ref{figuraC1}. 
We can nevertheless calculate the stringy volume of the $S^3$, at any value of the $\rho$-coordinate,
\begin{equation}
V_s[S^3]= \int \hat{\text{vol}}_{S^3}\, e^{-\Phi}\sqrt{\det[g+B]}=2\pi^2 u \sqrt{\frac{h_8}{ \hat{h}_4} }.
\end{equation}
This is finite at the ends of the $\rho$-interval. Hence, branes wrapping this $S^3$ will have finite action.

\end{document}